\documentclass[pra,10pt,twocolumn,aps,showpacs,superscriptaddress]{revtex4-1}
\usepackage{amsmath}
\usepackage{amssymb}
\usepackage{graphicx}
\usepackage{bm}
\usepackage[colorlinks=true,linkcolor=blue,citecolor=blue,urlcolor=blue]{hyperref}
\usepackage{txfonts}

\begin{document}
\title{Semiclassical excited-state signatures of quantum phase transitions\\ in spin chains with variable-range interactions}

\author{Manuel Gessner}
\email{manuel.gessner@ino.it}
\affiliation{Physikalisches Institut, Albert-Ludwigs-Universit\"at Freiburg, Hermann-Herder-Stra\ss e 3, 79104 Freiburg, Germany}
\affiliation{QSTAR (Quantum Science and Technology in Arcetri) and\\LENS (European Laboratory for Non-Linear Spectroscopy), Largo Enrico Fermi 2, I-50125 Firenze, Italy}
\affiliation{INRIM (Istituto Nazionale di Ricerca Metrologica), I-10135 Torino, Italy}
\author{Victor Manuel Bastidas}
\affiliation{Institut f\"ur Theoretische Physik, Technische Universit\"at Berlin, Hardenbergstra\ss e  36, 10623 Berlin, Germany}
\author{Tobias Brandes}
\affiliation{Institut f\"ur Theoretische Physik, Technische Universit\"at Berlin, Hardenbergstra\ss e  36, 10623 Berlin, Germany}
\author{Andreas Buchleitner}
\affiliation{Physikalisches Institut, Albert-Ludwigs-Universit\"at Freiburg, Hermann-Herder-Stra\ss e 3, 79104 Freiburg, Germany}

\date{\today}


\begin{abstract}
We study the excitation spectrum of a family of transverse-field spin chain models with variable interaction range and arbitrary spin $S$, which in the case of $S=1/2$ interpolates between the Lipkin-Meshkov-Glick and the Ising model. For any finite number $N$ of spins, a semiclassical energy manifold is derived in the large-$S$ limit employing bosonization methods, and its geometry is shown to determine not only the leading-order term but also the higher-order quantum fluctuations. Based on a multi-configurational mean-field ansatz, we obtain the semiclassical backbone of the quantum spectrum through the extremal points of a series of one-dimensional energy landscapes -- each one exhibiting a bifurcation when the external magnetic field drops below a threshold value. The obtained spectra become exact in the limit of vanishing or very strong external, transverse magnetic fields. Further analysis of the higher-order corrections in $1/\sqrt{2S}$ enables us to analytically study the dispersion relations of spin-wave excitations around the semiclassical energy levels. Within the same model, we are able to investigate quantum bifurcations, which occur in the semiclassical ($S\gg 1$) limit, and quantum phase transitions, which are observed in the thermodynamic ($N\rightarrow\infty$) limit.
\end{abstract}

\maketitle

\section{Introduction}
Phase transitions relate macroscopically observable, qualitative changes of the properties of a material to its microscopic structure and order. Thermally-driven phase transitions are usually described in terms of canonical ensembles, taking into account all the possible microstates of a system under appropriate boundary conditions \cite{sommerfeld,landau,reichl}. Since quantum phase transitions occur as a function of external control parameters at strictly zero temperature \cite{QPT1,Sachdev,QPT2}, when considering many-particle systems one is often tempted to restrict the theoretical treatment to a single quantum state -- the ground state. In fact, according to a widely employed definition \cite{Sachdev}, any non-analytic behavior of the ground state energy under smooth changes of the external control parameter in an infinitely extended lattice is considered a quantum phase transition. To understand the origin of such a non-analyticity, however, one has to consider excited states: Different eigenstates may -- due to their localization and/or symmetry properties -- respond differently to changes of the external parameter, which can cause an excited state to cross the ground state from above. In an adiabatic picture, this naturally leads to non-analytic behavior of the ground state energy and, thus, evokes the quantum phase transition. In the presence of non-vanishing couplings between the eigenstates, one instead observes avoided crossings, which generate jumps of the second derivative of the ground state energy  with respect to the control parameter, and for this reason in a many-particle context are called second-order quantum phase transitions.

Crossings and anti-crossings are however hardly specific to the ground state. Inasmuch as these are the expression of fundamental changes in the structural properties of the system -- as in the above classical understanding of phase transitions -- and not just the consequence of a perturbation-induced, local coupling between isolated pairs of states, one should in general expect dramatic structural changes throughout the entire spectrum, in the vicinity of a quantum phase transition. Indeed, quantum phase transitions are often accompanied \cite{Wu,Brandes03,kolovsky2004} by chaotic level statistics \cite{Haake,Mehta}, which can lead to rich dynamics in the vicinity of the critical point \cite{EPL}. Moreover, level clusterings in the excited states \cite{cusp,Heiss,ESQPT,ESQPT2} or in quasienergy states of driven systems \cite{Victor} have been identified as analogs of quantum phase transitions. However, a compelling general connection between the rearrangement of excited-state levels and the ground-state quantum phase transition is still lacking.

In this work we develop semiclassical methods, based on variational approaches and bosonization techniques, to study the excitation spectrum of spin chain models with tunable interaction range undergoing a quantum phase transition. To be able to investigate both, the effect of the finite interaction range, and the interplay of thermodynamic and semiclassical limits, we introduce a model of $N$ interacting spins whose respective length $S$ naturally defines an effective Planck constant as $\hbar_{\text{eff}}=1/S$. The semiclassical ($S\gg 1$) and thermodynamic ($N\rightarrow\infty$) limits of our model are fundamentally different---yet, non-analytic behavior can be observed in both cases. To distinguish between the two cases, we introduce the term \textit{quantum bifurcation}, which describes non-analyticities of the ground state energy of infinitely-connected, semiclassical (e.g. mean-field) models; see also~\cite{qb1,qb2,qb3}. Such quantum bifurcations are encountered in the semiclassical limit of our model even if $N$ is finite, whereas for finite $S$ we observe a quantum phase transition in the thermodynamic limit.

The central element of our analysis is a multi-dimensional semiclassical energy landscape, whose geometry directly determines the $1/\sqrt{2S}$-expansion of the Hamiltonian for large $S$. In the first part of this paper, we study this multi-dimensional energy landscape obtained in the semiclassical limit $S\gg 1$ for a finite number $N$ of spins. Imposing suitably chosen constraints, we obtain a series of one-dimensional sections of this energy landscape, whose extremal points reproduce key features of the full quantum spectrum, even in the most quantum case of $S=1/2$. The obtained semiclassical spectrum is furthermore shown to converge to the exact quantum spectrum when the external magnetic field -- the relevant control parameter -- is either very large or very small. In the second part, employing a bosonized representation of the spin algebra, we study the quantum fluctuations contained in terms of higher order in $1/\sqrt{2S}$. This allows us to identify the elementary spin-wave excitations of long-range interacting systems, and their dispersion relations. The disappearance of the excitation gap for the spin waves further predicts the exact critical point of the quantum bifurcation in a ring geometry with arbitrary $N$ and arbitrary interaction range. This critical point is shown to coincide with the bifurcation point of the corresponding semiclassical energy landscape. Increasing the number $N$ of spins, we observe the behavior of the spin waves close to the critical point as the system evolves from an effective semiclassical few-body system to an infinitely extended many-body system in the thermodynamic limit, $N\rightarrow\infty$, with tunable interaction range.

The model proposed and investigated in this paper includes, as special cases, several models that have been of recent experimental and theoretical interest. Examples include the long-range Ising model, which can be realized with strings of trapped ions for spin-$1/2$ systems \cite{PorrasCiracSpins,Schaetz,ExpSpinChains} and, recently, also for \mbox{spin-$1$} \cite{Senko}, as well as the conventional Ising model with nearest-neighbor interactions, which may be studied with cold atoms in tilted optical lattices \cite{Sachdev2,Simon,Meinert}.
Models with algebraically decaying, long-range interactions are able to account for the finite interaction length of, e.g., dipolar \cite{Rydbergs,PolarMolecules,Alvarez} or Coulomb \cite{PorrasCiracSpins} interactions, and allow to assess the modified spreading behaviour of perturbations when compared \cite{ExpSpinChains,Storch} to lattices with nearest-neighbor interactions \cite{LiebR,Cheneau}.

\section{The model}
\begin{figure}[b]
\centering
\includegraphics[width=.50\textwidth]{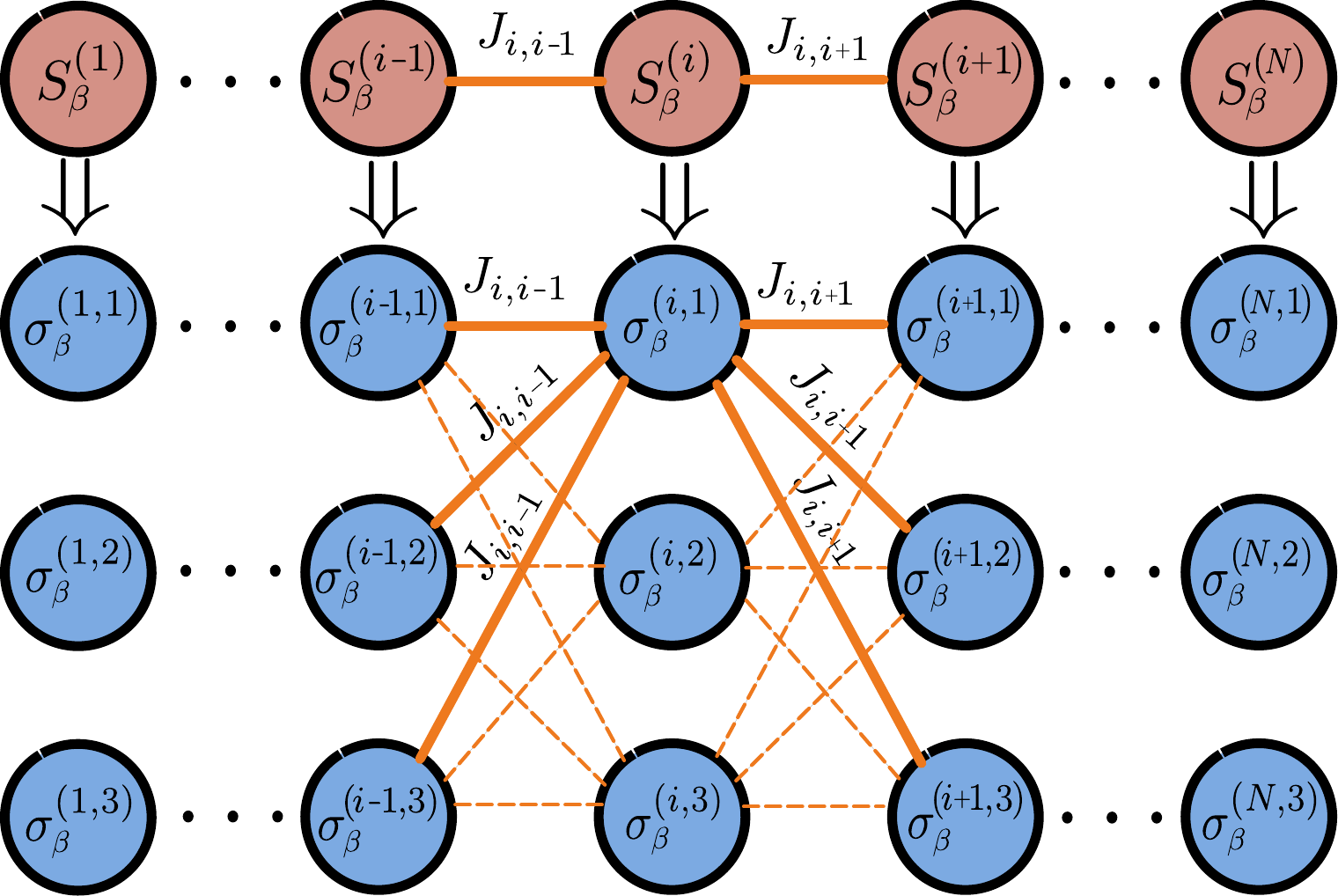}
\caption{Sketch of the model for a chain with $N$ sites and $M=3$. The upper part of the sketch depicts the spin chain in terms of collective angular momentum operators $S_{\beta}^{(i)}$ at the $i$-th site. Correspondingly, the lower part shows the representation in terms of the elementary spins $\sigma^{(i,k)}_{\beta}$ with $\beta\in\{x,y,z\}$ and $k=1,2,3$ . }
\label{fig.LatticeAdditional}
\end{figure}
We consider a one-dimensional variable-range spin model
\begin{align}\label{eq.spinchainS}
H = -\frac{2}{S}\sum_{\substack{i,j=1\\(i < j)}}^N J_{i,j} S_x^{(i)} S_x^{(j)} - 2B \sum_{i=1}^N S_y^{(i)},
\end{align}
where the spin-spin coupling reads $J_{i,j}=J_0/|i-j|^{\alpha}$. Furthermore, we have defined collective angular momentum operators $S_{\beta}^{(i)}=(1/2)\sum_{k=1}^{M}\sigma^{(i,k)}_{\beta}$ with $\beta\in\{x,y,z\}$ such that $[S_{x}^{(i)},S_{y}^{(j)}]=\mathrm{i}\delta_{ij}S_{z}^{(j)}$. In addition, $\sigma^{(i,k)}_{\beta}$ for $k=1,\dots,M$ are Pauli matrices describing $M$ elementary spins at the $i$-th site, in such a way that $S=M/2$.  In the course of this paper we will discuss both cases of open and periodic boundary conditions. Figure~\ref{fig.LatticeAdditional} depicts a sketch of the model and its interpretation in terms of collective angular momentum operators $S_{\beta}^{(i)}$ and elementary spins $\sigma^{(i,k)}_{\beta}$. In the special case of $S=1/2$ \cite{Cannas,CannasRenor,Cardy,Dyson,PorrasCiracSpins,Schaetz,ExpSpinChains,Koffel,EPL,PhDGessner}, the Hamiltonian~\eqref{eq.spinchainS} reads
\begin{align}\label{eq.spinchain}
H = -J_0\sum_{\substack{i,j=1\\(i < j)}}^N \frac{1}{|i-j|^\alpha}\sigma_x^{(i)} \sigma_x^{(j)} - B \sum_{i=1}^N \sigma_y^{(i)}.
\end{align}
As a function of $\alpha$, which determines the interaction range, the Hamiltonian~\eqref{eq.spinchain} interpolates continuously between the infinite-range Lipkin-Meshkov-Glick model ($\alpha=0$) \cite{LMG} and the one-dimensional Ising model with nearest-neighbor interactions ($\alpha=\infty$) \cite{Ising,Sachdev}. 

The system's properties are determined by the relative strength of the two competing interactions: The internal spin-spin interaction $J_0$ causes the spins to arrange their configuration depending on the $x$-coordinates of neighboring spins, while the external field $B$ pushes the spins along the transverse $y$-direction. The sign of $J_0$ determines whether the system arranges in ferromagnetic ($J_0>0$) or \mbox{(anti-)}ferro\-magnetic ($J_0<0$) order in the limit $B=0$. The quantum phase transition occurs when the two potential energy terms proportional to $B$ and $J_0$ are of comparable order of magnitude, whereas the exact position of the critical point depends on $\alpha$.

For $S=1/2$, the phase transition has been studied for the special cases of the Ising and Lipkin-Meshkov-Glick models -- analytic solutions are available for both of them \cite{Lieb,Pan}. For \mbox{$\alpha=\infty$} the system can be solved by Jordan-Wigner fermionization \cite{JordanWigner}, and exhibits a quantum phase transition from \mbox{(anti-)}ferro\-magnet to paramagnet at the critical field \mbox{$B_c=|J_0|$}~\cite{Sachdev}. In the opposite limit $\alpha=0$, a Holstein-Primakoff bosonization \cite{HP} yields an efficient description of the system in orders of $1/N$ (since all spins can be combined into one large spin) \cite{Dusuel}, which is more practical than its exact solution \cite{LMGVidal}. Whenever $\alpha\leq 1$, the spectrum is only bounded in the thermodynamic limit when $J_0$ is rescaled by $N$ \cite{Cannas}. The quantum bifurcation of the Lipkin-Meshkov-Glick model ($\alpha=0$) occurs at $B_c=\bar{J}_0$ with $\bar{J}_0=J_0/N$ when $J_0>0$ \cite{Botet,Zibold}, and at $B=0$ when $J_0<0$ \cite{Vidal}. Only few results are available for intermediate values of $\alpha$ \cite{Koffel}. A phase transition in a classical long-range model for $S=1/2$ was shown to occur for the parameter range $1<\alpha<2$ \cite{Dyson}. Further studies based on renormalization group techniques allowed to investigate the particular case $\alpha=2$ \cite{Cardy} and to describe non-analyticities of the free energy as a function of $\alpha$ \cite{CannasRenor}.

Conversely, in the semiclassical limit of Eq.~(\ref{eq.spinchainS}), $S\gg 1$, the number $M=2S$ of elementary spins at each lattice site becomes very large. As a consequence, each site $i$ is represented by a composite semiclassical spin, as depicted in Fig.~\ref{fig.LatticeAdditional}, 
while different sites are coupled by a finite interaction range, determined by $\alpha$. 
The semiclassical limit allows for an exact mean-field analysis and produces sharp bifurcations of the energy landscapes, which directly imply non-analytic behavior of the quantum excitations for all values of $N\geq 2$. These phenomena are henceforth referred to as quantum bifurcations, to distinguish them from the quantum phase transition in systems that are infinitely extended along the interacting dimension. A discussion of quantum bifurcations will be provided in Sec.~\ref{sec.QuantumBifurcationbosons}, where the main features are illustrated with a simple special case of our model. A complete analysis of the quantum bifurcations in our model is then provided in Sec.~\ref{sec.spinwaves}. The model additionally permits us to tune the number $N$ of composite spins to independently scan the transition to the thermodynamic limit, which is associated with the transition into an infinitely extended one-dimensional lattice whose interaction range is parametrized by $\alpha$.


\section{Semiclassical expansion of the spin Hamiltonian}\label{sec.bosons}
We begin by deriving a formal semiclassical expansion ($S\gg 1$) of the Hamiltonian~(\ref{eq.spinchainS}) for a finite number $N$ of spins. In this limit, the spectrum of each individual spin resembles that of a harmonic oscillator, which allows us to express the spin operators in terms of bosonic creation and annihilation operators. The associated Holstein-Primakoff transformation \cite{HP} then leads to a perturbative expansion in orders of $1/\sqrt{2S}$.

\subsection{General formalism}\label{sec.Generbosons}
To obtain the semiclassical expansion, we restrict ourselves to the subspace of maximal angular momentum $S=M/2$. In addition, it is convenient to introduce a local rotation operator of the $i$-th spin as
\begin{align}
U(\phi^{(i)})=\exp(\mathrm{i}\phi^{(i)}S^{(i)}_z).
\label{eq:UnitCoherentState}
\end{align}
Based on these local spin rotations, we introduce the rotated Hamiltonian as in Ref.~\cite{Dusuel}
\begin{align}
H(\boldsymbol{\phi})=U(\boldsymbol{\phi})HU^{\dagger}(\boldsymbol{\phi}),
\end{align}
which is given by
\begin{align}
      \label{eq:RotHam}
            H(\boldsymbol{\phi})
            &=-\frac{2J_0}{S}\sum_{\substack{i,j=1\\(i < j)}}^N \frac{1}{|i-j|^\alpha}\left[S_x^{(i)} S_x^{(j)}\cos\phi^{(i)}\cos\phi^{(j)}\right.\notag\\&\hspace{2.7cm}-S_x^{(i)} S_y^{(j)}\cos\phi^{(i)}\sin\phi^{(j)}\notag\\&\hspace{2.7cm}-S_y^{(i)} S_x^{(j)}\sin\phi^{(i)}\cos\phi^{(j)}\notag\\&\hspace{2.7cm}\left.+S_y^{(i)} S_y^{(j)}\sin\phi^{(i)}\sin\phi^{(j)}\right]
            \notag\\&\quad
            - 2B \sum_{i=1}^N \left[S_y^{(i)}\cos\phi^{(i)}+S_x^{(i)}\sin\phi^{(i)}\right].
\end{align}
In this expression, $U(\boldsymbol{\phi})=\bigotimes^{N}_{i=1}U(\phi^{(i)})$ is a tensor product of the unitary operators defined in Eq.~\eqref{eq:UnitCoherentState}, and the local spin orientations are characterized by the vector $\boldsymbol{\phi}=(\phi^{(1)},\dots,\phi^{(N)})$. Since $U(\boldsymbol{\phi})$ represents a unitary operation, the spectra of $H$ and $H(\boldsymbol{\phi})$ coincide. We now invoke the Holstein-Primakoff representation of the angular momentum algebra~\cite{HP}:
\begin{align}
	    S^{(i)}_y &=S-a_{i}^{\dagger}a_{i} ,\label{HPZ} \\ 
	    S^{(i)}_{x}+\mathrm{i}S^{(i)}_{z}&=\sqrt{2S}a_{i}^{\dagger}~\left(1-\frac{a_{i}^{\dagger}a_{i}}{2S}\right)^{1/2} ,\label{HPMas} \\ 
            S^{(i)}_{x}-\mathrm{i}S^{(i)}_{z}&=\sqrt{2S}\left(1-\frac{a_{i}^{\dagger}a_{i}}{2S}\right)^{1/2}~a_{i}, \label{HPMen}
\end{align}
where $S$ is the total angular momentum at the $i$-th site and $a_{i},a_{i}^{\dagger}$ are bosonic operators. In the case $S\gg1$ one can expand the Hamiltonian~\eqref{eq:RotHam} as
\begin{align}
      \label{eq:HamTaylorS}
            H(\boldsymbol{\phi})=2SN E(\boldsymbol{\phi})+\sqrt{2S}H_{\mathrm{L}}(\boldsymbol{\phi})+H_{\mathrm{Q}}(\boldsymbol{\phi})+\mathcal{O}\left(\frac{1}{\sqrt{2S}}\right),
\end{align}
where the leading-order term in $S$ defines a semiclassical energy landscape
\begin{align}
      \label{eq:HamTaylorEnergy}
            E(\boldsymbol{\phi})=-\frac{J_0}{N}\sum_{\substack{i,j=1\\(i < j)}}^N \frac{\sin\phi^{(i)}\sin\phi^{(j)}}{|i-j|^\alpha}- \frac{B}{N} \sum_{i=1}^N \cos\phi^{(i)}.
\end{align}
In the derivation of the energy landscape we considered the case of maximal angular momentum. However, in Section~\ref{sec.onedim} we show how this assumption can be relaxed to obtain a generalized energy landscape by using a variational approach.

The partial derivatives of $E(\boldsymbol{\phi})$ further determine the linear Hamiltonian $H^{\mathrm{L}}(\boldsymbol{\phi})$, containing quantum corrections in linear order,
\begin{align}\label{eq.Hlin}
    H_{\mathrm{L}}(\boldsymbol{\phi})  &=-N\sum_{i=1}^N\frac{\partial E(\boldsymbol{\phi})}{\partial \phi^{(i)}}(a^{\dagger}_i+a_i).
\end{align}
as well as the quadratic Hamiltonian,
\begin{align}\label{eq.Hquad}
 H_{\mathrm{Q}}(\boldsymbol{\phi})&=N\sum_{\substack{i,j=1\\(i < j)}}^N \frac{\partial^2 E(\boldsymbol{\phi})}{\partial \phi^{(i)}\partial \phi^{(j)}}(a_{i}^{\dagger}+a_{i})(a_{j}^{\dagger}+a_{j})\notag\\&\quad +2N \sum_{i=1}^N \frac{\partial^2 E(\boldsymbol{\phi})}{\partial \phi^{(i)2}}a_{i}^{\dagger}a_{i}.
\end{align}

Thus, we find that the semiclassical energy landscape $E(\boldsymbol{\phi})$ contains the complete information about the Hamiltonian for large $S$, and determines, via its geometry, also the quantum corrections to the mean-field contribution. Of special interest are the stationary points $\mathcal{C}$ of $E(\boldsymbol{\phi})$, which are defined as those spin configurations $\boldsymbol{\phi}$ that satisfy the conditions
\begin{align}
      \label{eq:HamTaylorFixedPoints}
            \frac{\partial E(\boldsymbol{\phi})}{\partial \phi^{(i)}}=-\frac{J_0}{N}\sum_{\substack{j=1\\(i \neq j)}}^N \frac{\cos\phi^{(i)}\sin\phi^{(j)}}{|i-j|^\alpha}+\frac{B}{N} \sin\phi^{(i)}=0,
\end{align}
for all $i=1,\dots,N$. At these points, the linear Hamiltonian $H^{\mathrm{L}}(\boldsymbol{\phi})$ vanishes exactly, and the quantum fluctuations are described by $H^{\mathrm{Q}}(\boldsymbol{\phi})$, whose coefficients are given by
\begin{align}\label{eq.diagsecondderv}
\frac{\partial^2 E(\boldsymbol{\phi})}{\partial \phi^{(i)2}}=\frac{J_0}{N}\sum_{\substack{j=1\\(i \neq j)}}^N \frac{\sin\phi^{(i)}\sin\phi^{(j)}}{|i-j|^\alpha}+\frac{B}{N} \cos\phi^{(i)},
\end{align}
and, for $i\neq j$,
\begin{align}
 \frac{\partial^2 E(\boldsymbol{\phi})}{\partial \phi^{(i)}\partial \phi^{(j)}}=-\frac{J_0}{N} \frac{\cos\phi^{(i)}\cos\phi^{(j)}}{|i-j|^\alpha}.
 \end{align}
Using Eq.~(\ref{eq:HamTaylorFixedPoints}) in Eq.~(\ref{eq.diagsecondderv}) simplifies the second derivative at a stationary point to
\begin{align}\label{eq.simpldiagsecder}
\left.\frac{\partial^2 E(\boldsymbol{\phi})}{\partial \phi^{(i)2}}\right|_{\boldsymbol{\phi}\in\mathcal{C}}=\frac{B}{N}\frac{1}{\cos\phi^{(i)}},
\end{align}
provided that $\cos\phi^{(i)}\neq 0$.

Section~\ref{eq.semiclassicallandscapes} is dedicated to an analysis of the zero-order semiclassical energy landscape $E(\boldsymbol{\phi})$. In particular we will compare the semiclassical predictions for the energy spectrum based on a series of suitably defined one-dimensional sections of $E(\boldsymbol{\phi})$ to the numerically obtained quantum spectra, with particular emphasis on the least classical case of $S=1/2$. In this parameter regime far away from the semiclassical limit, the quantum phase transition can be observed in the thermodynamic limit when the external field $B$ approaches a critical value that depends on $\alpha$. 

Later in Section~\ref{sec.spinwaves} we study the quantum fluctuations in the large-$S$ limit, which in turn allows us to reveal quantum bifurcations for arbitrary $N$ and $\alpha$ by analytical means. We conclude the present section by formally introducing the concept of quantum bifurcations, based on a simple illustrative example, and by discussing its relation to quantum phase transitions. 

\subsection{Quantum bifurcations in the semiclassical limit}\label{sec.QuantumBifurcationbosons}
In this section, we discuss the Hamiltonian~\eqref{eq:HamTaylorS} for $J_0>0$ in the particular case of just two lattice sites $N=2$ to illustrate the main features of quantum bifurcations. For a complete analysis of the quantum bifurcations in the model and additional details on the employed methods, we refer to Sec.~\ref{sec.spinwaves}. Let us further assume an equal spin configuration $\boldsymbol{\phi}_c=(\phi_c,\phi_c)$, where $\phi_c$ is chosen such that the mean-field energy $E(\boldsymbol{\phi}_c)$ is minimized. We find a single minimum at $\phi_c=0$ when $B\geq J_0$ and two degenerate minima at $\phi_c=\pm \arccos(B/J_0)$ when $B<J_0$. As we will see in this section, this bifurcation of the classical mean-field energy landscape entails profound consequences for the quantum fluctuations of the higher-order terms in Eq.~(\ref{eq:HamTaylorS}).

Due to Eq.~\eqref{eq:HamTaylorFixedPoints}, the linear Hamiltonian $H_{\mathrm{L}}(\boldsymbol{\phi})$ disappears at critical points of $E(\boldsymbol{\phi})$. The Hamiltonian~(\ref{eq:HamTaylorS}) thus reduces to the mean-field energy and the quadratic corrections. Through their dependence on $\boldsymbol{\phi}_c$, both terms depend on the parameter $B/J_0$. For $B\geq J_0$, i.e., $\phi_c=0$, we obtain the effective Hamiltonian 
\begin{align}
\label{eq:QuantumBifurcation1}
H(\boldsymbol{\phi}_c)=-4SB-J_0(a^{\dagger}_{1}+a_{1})(a^{\dagger}_{2}+a_{2})+2B(a^{\dagger}_{1}a_{1}+a^{\dagger}_{2}a_{2}).
\end{align}
This Hamiltonian, in fact, can be identified with the effective Hamiltonian for the on-resonance Dicke model \cite{dicke} in the normal phase, 
where $2B$ plays the role of atomic and mode resonances, $-J_0$ reflects the atom-field coupling strength, and $2S$ is the collective atomic spin \cite{Brandes03}. A diagonalization of the Hamiltonian~(\ref{eq:QuantumBifurcation1}) leads to two collective bosonic modes with excitation energies $\varepsilon_{\pm}=2\sqrt{B(B\pm J_0)}$. Correspondingly, $\varepsilon_{-}$ is the gap between the ground state and the first excited state.

When $B<J_0$, i.e., $\cos\phi_c=B/J_0$, the Hamiltonian reads
\begin{align}
\label{eq:QuantumBifurcation2}
H(\boldsymbol{\phi}_c)&=-2S(J_0+B^2/J_0)-B^2/J_0(a^{\dagger}_{1}+a_{1})(a^{\dagger}_{2}+a_{2})\notag\\
&\quad+2J_0(a^{\dagger}_{1}a_{1}+a^{\dagger}_{2}a_{2}),
\end{align}
Now, we obtain collective excitation energies of $\varepsilon_{\pm}=2\sqrt{J_0^2\pm B^2}$.

The above results indicate that when $B\rightarrow J_{0}$ the energy gap above the ground state vanishes
as $\varepsilon_{-}\sim|B-B_{\text{c}}|^{1/2}$, where $B_{c}=J_0$ is the bifurcation point. The energy gap directly defines a characteristic length scale $l_{-}=1/\sqrt{\varepsilon_{-}}$ which determines the spread of the ground state wave function~\cite{Brandes03}. Indeed, the ground state, which is a two-mode Gaussian state, gets strongly squeezed as the 
bifurcation point is approached~\cite{Brandes03}. Similarly, the ground state shows strong quantum correlations in the vicinity of the bifurcation point \cite{Brandes04}. Furthermore, the divergent length scale $l_{-}$ and the closing gap $\varepsilon_{-}$ can be associated with critical exponents, and finite-size scaling can be studied when $S$ is finite \cite{Brandes03,Dusuel,LMGVidal,Botet,Brandes04}. In the semiclassical limit, we find a sharp discontinuity of the ground state energy at the bifurcation point.

Evidently, these quantum signatures of the classical bifurcation stand in direct analogy to quantum phase transitions. The quantum phase transition, however, occurs in the thermodynamic limit, when the extension of the lattice of interacting, collective spins becomes infinite. In the semiclassical limit, which triggers the quantum bifurcation for any value of $N$, we extend the zero-dimensional sub-lattice of elementary spins, which add up to form a collective spin as it is depicted in Fig.~\ref{fig.LatticeAdditional}. According to Eqs.~\eqref{HPZ}, \eqref{HPMas} and \eqref{HPMen}, the number $M$ of elementary spins translates into the maximal occupation of the effective bosonic modes, and consequently, only in the semiclassical limit, these modes are unbounded and allow for a diverging spread of the ground state wavefunction. This way, the many-body character of the spin model is absorbed by a finite number $N$ of harmonic oscillator modes. The diverging length scale, however, despite being related to the ground-state correlations, does not identify a diverging \textit{spatial} correlation length, since the elementary spins are arranged on a zero-dimensional lattice (see Fig.~\ref{fig.LatticeAdditional}). This precisely identifies the difference between 
the semiclassical and thermodynamic limits, i.e., the quantum bifurcation and the quantum phase transition, respectively.

A further characteristic of the quantum bifurcation is that the mean-field description becomes exact in the classical limit $S\rightarrow\infty$. Moreover, the many-body aspect of the model for finite $N$ becomes irrelevant in the semiclassical limit: The quantum bifurcation does not explicitly depend on the substructure of the collective spins. The features close to the ground state are therefore reproduced by an effective $N$-body system rather than an $MN$-body system. Yet, for an understanding of the excitation spectrum, this substructure is essential, as will become apparent in Sec.~\ref{eq.semiclassicallandscapes}.

\section{Semiclassical energy landscapes}\label{eq.semiclassicallandscapes}
The semiclassical energy landscape $E(\boldsymbol{\phi})$ was derived as the leading-order contribution to a $1/\sqrt{2S}$-expansion in a subspace of maximal angular momentum. We will show in the first part of this Section that this term can also be interpreted as the energy expectation value of a variational product ansatz of spin-coherent states in this particular subspace. Spin coherent states are formal analogues of coherent states of the harmonic oscillator \cite{Radcliffe,Arecchi,Zhang,MandelWolf}. Their expectation values are characterized by Bloch vector coordinates, and they minimize the uncertainty relation with respect to certain angular momentum observables \cite{Arecchi}. These states are therefore often interpreted as semiclassical, and they allow to introduce an effective Planck constant $\hbar_{\mathrm{eff}}=1/S$ \cite{Haake,Zhang}. 

Spin-coherent states are, however, not limited to the subspace of maximal angular momentum $S$. We can therefore formulate a more general variational ansatz in terms of arbitrary spin-coherent states that will allow us to extend the definition of the semiclassical energy landscape~(\ref{eq:HamTaylorEnergy}) to include all subspaces of $S$.

\subsection{Variational approach based on spin coherent states}\label{sec.onedim}
In this section we employ a product state ansatz for the trial wave function in terms of local spin coherent states
\begin{align}\label{eq.trialstate}
|\mathbf{l}_{\mu},\boldsymbol{\phi}_{\mu}\rangle=\bigotimes_{i=1}^N|l_{\mu}^{(i)},\phi_{\mu}^{(i)}\rangle.
\end{align}
The trial states~(\ref{eq.trialstate}) are characterized by the vectors $\mathbf{l}_{\mu}=(l_{\mu}^{(1)},l_{\mu}^{(2)},\dots,l_{\mu}^{(N)})$ and $\boldsymbol{\phi}_{\mu}=(\phi_{\mu}^{(1)},\phi_{\mu}^{(2)},\dots,\phi_{\mu}^{(N)})$, which depend on the configuration labelled by $\mu$. The variables $l_{\mu}^{(i)}$ and $\phi_{\mu}^{(i)}$ determine the respective length and orientation of the local spin coherent state which describes the spin at index $i$ as 
\cite{MandelWolf}
\begin{align}\label{eq.scstate}
|l_{\mu}^{(i)},\phi_{\mu}^{(i)}\rangle=\frac{1}{2^{l_{\mu}^{(i)}}}\sum_{m_z^{(i)}=-l_{\mu}^{(i)}}^{l_{\mu}^{(i)}}\binom{2l_{\mu}^{(i)}}{l_{\mu}^{(i)}+m_z^{(i)}}^{1/2}\left(-ie^{i\phi_{\mu}^{(i)}}\right)^{l_{\mu}^{(i)}+m_z^{(i)}}|l_{\mu}^{(i)},m_z^{(i)}\rangle.
\end{align}
Here $|l_{\mu}^{(i)},m_z^{(i)}\rangle$ are the Dicke states \cite{dicke} of cooperation number $0\leq l_{\mu}^{(i)}\leq M/2$, as defined by a total angular momentum of $S^{(i)2}|l_{\mu}^{(i)},m_z^{(i)}\rangle=l_{\mu}^{(i)}(l_{\mu}^{(i)}+1)|l_{\mu}^{(i)},m_z^{(i)}\rangle$, and $S^{(i)}_z|l_{\mu}^{(i)},m_z^{(i)}\rangle=m_z^{(i)}|l_{\mu}^{(i)},m_z^{(i)}\rangle$. 

This ansatz, by construction, ensures that the local expectation values are restricted to the $xy$-plane. This choice is motivated by the fact that $S^{(i)}_z$ does not appear in the Hamiltonian~(\ref{eq.spinchainS}), and therefore does not contribute to the energy. The expectation values are now conveniently represented by the Bloch vector coordinates as:
\begin{align}
\langle l_{\mu}^{(i)},\phi_{\mu}^{(i)}|S^{(i)}_x|l_{\mu}^{(i)},\phi_{\mu}^{(i)}\rangle&=l_{\mu}^{(i)}\sin\phi_{\mu}^{(i)},\label{eq.exp1}\\
\langle l_{\mu}^{(i)},\phi_{\mu}^{(i)}|S^{(i)}_y|l_{\mu}^{(i)},\phi_{\mu}^{(i)}\rangle&=l_{\mu}^{(i)}\cos\phi_{\mu}^{(i)},\label{eq.exp2}\\
\langle l_{\mu}^{(i)},\phi_{\mu}^{(i)}|S^{(i)}_z|l_{\mu}^{(i)},\phi_{\mu}^{(i)}\rangle&=0.\label{eq.exp3}
\end{align}

Each of the $N$ spins is composed of $M$ spin-$1/2$ particles, leading to a total of $MN=2SN$ elementary spins in the system. Using the trial states~(\ref{eq.trialstate}), together with (\ref{eq.spinchainS},\ref{eq.exp1},\ref{eq.exp2}), we obtain the average energy per elementary spin
\begin{align}\label{eq.averageenergygeneral}
&\quad E_{\mu}(\mathbf{l}_{\mu},\boldsymbol{\phi}_{\mu})=\frac{1}{2SN}\langle \mathbf{l}_{\mu},\boldsymbol{\phi}_{\mu}|H|\mathbf{l}_{\mu},\boldsymbol{\phi}_{\mu}\rangle\notag\\
&=-\frac{J_0}{S^2N}\sum_{\substack{i,j=1\\(i < j)}}^N \frac{l_{\mu}^{(i)}\sin\phi_{\mu}^{(i)}l_{\mu}^{(j)}\sin\phi_{\mu}^{(j)}}{|i-j|^\alpha}-\frac{B}{SN} \sum_{i=1}^N l_{\mu}^{(i)}\cos\phi_{\mu}^{(i)},
\end{align}
This generalized semiclassical energy landscape indeed coincides with the energy landscape~(\ref{eq:HamTaylorEnergy}) when we restrict to the subspace where all of the $N$ composite spins have maximal angular momentum $S$, i.e., when $l_{\mu}^{(i)}\equiv S$ for all $i=1,\dots,N$. In fact, the rotations introduced in Eq.~(\ref{eq:UnitCoherentState}) can be used to generate spin coherent states $|S,\phi^{(i)}\rangle$ in this particular subspace as $|S,\phi^{(i)}\rangle=U(\phi^{(i)})|S,m_y^{(i)}=S\rangle$, where $S_y^{(i)}|S,m_y^{(i)}\rangle=m_y^{(i)}|S,m_y^{(i)}\rangle$.

In the following we will show that, in the limiting cases $J_0=0$ and $B=0$, the ansatz~(\ref{eq.trialstate}) allows to generate the exact spectra of (\ref{eq.spinchainS}).

\subsection{(Anti-)ferromagnetic spectra at $B=0$}\label{eq.anitferro}
Let us first consider the spectrum in absence of an external field, $B=0$. We define the local eigenstates $|S,m^{(i)}_x\rangle$ by the eigenvalue equation $S_x^{(i)}|S,m^{(i)}_x\rangle=m^{(i)}_x|S,m^{(i)}_x\rangle$, with $m^{(i)}_x=-S,\dots,S$. Introducing
\begin{align}
|S,\mathbf{m}_x\rangle=\bigotimes_{i=1}^N|S,m^{(i)}_x\rangle,
\end{align}
with $\mathbf{m}_x=(m_x^{(1)},\dots,m_x^{(N)})$, the Hamiltonian describing the internal spin-spin interaction
\begin{align}
\label{eq.B0Ham}
H_{\mathrm{in}} =\left. H\right|_{B=0}= -\frac{2J_0}{S}\sum_{\substack{i,j=1\\(i < j)}}^N \frac{1}{|i-j|^\alpha}S_x^{(i)} S_x^{(j)},
\end{align}
which is obtained from (\ref{eq.spinchainS}) for $B=0$, satisfies the eigenvalue equation
\begin{align}\label{eq.exactferrospectrum}
H_{\mathrm{in}}|S,\mathbf{m}_x\rangle=-\frac{2J_0}{S}\sum_{\substack{i,j=1\\(i < j)}}^N \frac{m_x^{(i)} m_x^{(j)}}{|i-j|^\alpha}|S,\mathbf{m}_x\rangle.
\end{align}

To reproduce this exact spectrum~(\ref{eq.exactferrospectrum}) from the energy expectation value per spin~(\ref{eq.averageenergygeneral}), we impose certain conditions on the spin coherent state ansatz~(\ref{eq.trialstate}). In particular, we assume that all the local spin orientations $\phi^{(i)}_{\mu}$ are represented by the same angle $\phi$, while allowing individual spins to be \textit{inverted} such that $\phi_{\mu}^{(i)}=\pm\phi$. Thus, we see from Eq.~(\ref{eq.averageenergygeneral}) at $B=0$ that configurations with arbitrary sequences of positive or negative local angles $+\phi$ or $-\phi$, lead, at the particular value of $\phi=\pi/2$ (see below), to the scaled energy expectation value
\begin{align}\label{eq.expscstsb0}
E_{\mu}(\mathbf{l}_{\mu},\boldsymbol{\phi}_{\mu_{\mathrm{c}}})=-\frac{J_0}{S^2N}\sum_{\substack{i,j=1\\(i < j)}}^N \frac{l_{\mu}^{(i)}\epsilon^{(i)}_{\mu}l_{\mu}^{(j)}\epsilon^{(j)}_{\mu}}{|i-j|^{\alpha}},
\end{align}
where we have defined $\epsilon^{(i)}_{\mu}=\pm 1$ via the equality $\sin\phi_{\mu}^{(i)}=\epsilon^{(i)}_{\mu}\sin\phi$. The possible values of Eq.~(\ref{eq.expscstsb0}) can now be determined by scanning over the full range of $-S\leq l_{\mu}^{(i)}\epsilon^{(i)}_{\mu}\leq S$. Comparison to Eq.~(\ref{eq.exactferrospectrum}), while recalling that $-S\leq m^{(i)}_x\leq S$, demonstrates that the product state ansatz (\ref{eq.trialstate}) of trial states is able to reproduce the exact spectrum at $B=0$, i.e., the set of eigenvalues coincides with the set of all possible values of
\begin{align}\label{eq.totE}
\mathsf{E}_{\mu}=2SNE_{\mu}\, ,
\end{align}
where $\mathsf{E}_{\mu}$ represents the total energy, in contrast to the energy $E_{\mu}$ per elementary spin unit. Note that this observation is, in fact, independent of the coupling coefficients, which here are given by the algebraically decaying function $J_{i,j}=J_0/|i-j|^{\alpha}$.

\begin{figure}[tb]
\centering
\includegraphics[width=.49\textwidth]{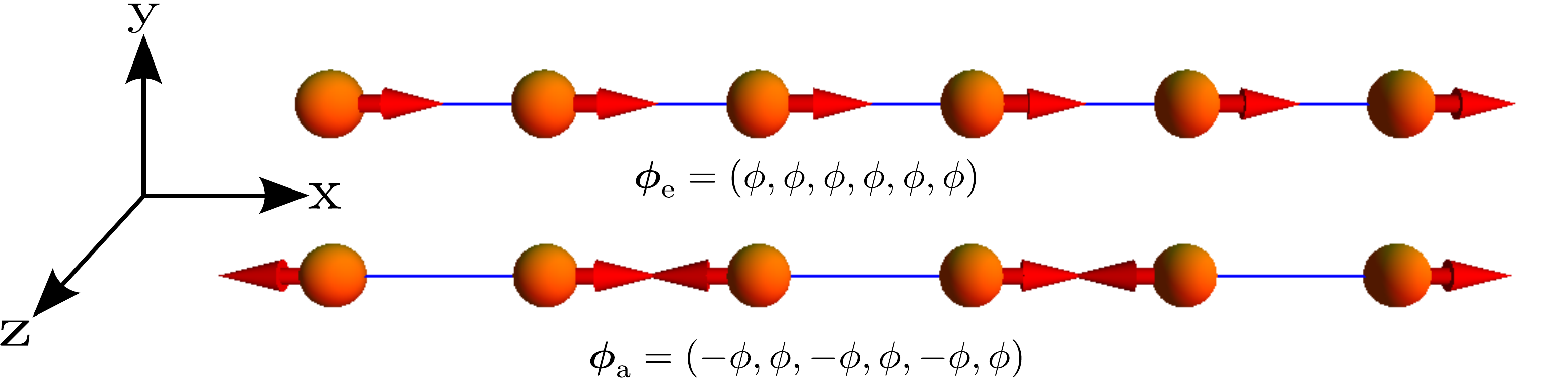}
\caption{
Spin configurations $\boldsymbol{\phi}_{\mathrm{e}}$ and $\boldsymbol{\phi}_{\mathrm{a}}$ associated with the largest and the smallest energy eigenvalue of (\ref{eq.B0Ham}), for $N=6$, $S=1/2$, and $\phi=\pi/2$. The spin coherent state (top), where all spins align in equal directions, generates the semiclassical ground state configuration when $J_0>0$. In this case, the alternating configuration (bottom) corresponds to the semiclassical configuration which yields the largest energy eigenvalue. In the special case of the Ising model ($\alpha\rightarrow\infty$), the configurations $\boldsymbol{\phi}_{\mathrm{e}}$ (equal directions) and $\boldsymbol{\phi}_{\mathrm{a}}$ (alternating directions) yield energies $E_{\mathrm{e}}=-J_0(N-1)$ and $E_{\mathrm{a}}=J_0(N-1)$, respectively [see Eq.~(\ref{eq.Isingspectrum})].}
\label{fig.ConfigSpin}
\end{figure}

Figure~\ref{fig.ConfigSpin} illustrates two examples of spin configurations $\boldsymbol{\phi}^{\mathrm{c}}_{\mu}$ for $N=6$, which differ by the number of inverted spins. The natural orientation of the spin-spin interaction along the $x$-direction causes the spins to assume an orientation along the $x$-axis in absence of the transverse field. Hence, the angle assumes the value $\phi=\pi/2$, see Eqs.~(\ref{eq.exp1})-(\ref{eq.exp3}). 

In total there are $2^N$ different configurations labeled by $\mu$. In the case $S=1/2$, this number indeed reflects the Hilbert space dimension. However, two symmetries lead to degeneracies of the $\mathsf{E}_{\mu}$: 
\begin{itemize}
\item[(i)] The invariance of the energy expectation value under a global sign flip, $\phi\rightarrow-\phi$, originates in the $Z_2$-symmetry~\cite{Sachdev,Brandes03} ($\pi$-rotation around the $y$-axis) of the Hamiltonian~(\ref{eq.spinchainS}) and permits to restrict our analysis to configurations with at most half of the spins inverted. 
\item [(ii)] For open boundary conditions, which we impose in this Section, the energy of the chain remains invariant under a mirror reflection with respect to the center: $i\rightarrow N-i+1$.
\end{itemize}

In the following we discuss the quantity~(\ref{eq.expscstsb0}) where, for simplicity, we focus on the special case of $S=1/2$. The analysis can be extended easily to larger spins. When $S=1/2$, the length of the individual spin coherent states is fixed at $l_{\mu}^{(i)}=1/2$, for all $i=1,\dots, N$. Any given configuration $(\mathbf{l}_{\mu},\boldsymbol{\phi}_{\mu})$ is then fully determined by the orientations $\boldsymbol{\phi}_{\mu}$ of the local spins. For this special case, we introduce the effective spin-spin coupling constant 
\begin{align}\label{eq.jeff}
\mathsf{J}_{\mu}=\frac{J_0}{N}\sum_{\substack{i,j=1\\(i < j)}}^N\frac{\epsilon^{(i)}_{\mu}\epsilon^{(j)}_{\mu}}{|i-j|^{\alpha}},
\end{align}
which determines the energy spectrum $\mathsf{E}_{\mu}$ at $B=0$ through $\mathsf{E}_{\mu}=NE_{\mu}(\boldsymbol{\phi}^{\mathrm{c}}_{\mu})=-N\mathsf{J}_{\mu}$ [where we used $S=1/2$ in Eqs.~(\ref{eq.expscstsb0},\ref{eq.totE})].

For arbitrary values of the interaction decay constant $\alpha$, the different possible configurations $\boldsymbol{\phi}^{\mathrm{c}}_{\mu}$ lead to rather irregular distributions of the energy eigenvalues. However, in certain extreme cases, when we recover Ising ($\alpha=\infty$) or Lipkin-Meshkov-Glick interactions ($\alpha=0$), only few, strongly degenerate energy bands are obtained. 

For example, in the limit of nearest-neighbor interactions, $\alpha=\infty$, we infer from (\ref{eq.jeff})
\begin{align}\label{eq.Isingspectrum}
\mathsf{E}_{\mu}=-J_0\sum_{i=1}^{N-1}\epsilon_{\mu}^{(i)}\epsilon_{\mu}^{(i+1)}=-J_0(N-1-2r_{\mu}),
\end{align}
where $r_{\mu}=0,\dots,N-1$ counts the number of \textit{domain walls} in the configuration $\boldsymbol{\phi}_{\mu}$. In a ferromagnet (anti-ferromagnet), these occur when two neighboring spins align in opposite (equal) directions \cite{Sachdev}. 

Conversely, for an infinitely extended interaction range, $\alpha=0$, we have
\begin{align}\label{eq.LMGspectrum}
\mathsf{E}_{\mu}=-J_0\sum_{\substack{i,j=1\\ (i\neq j)}}^N\epsilon_{\mu}^{(i)}\epsilon_{\mu}^{(j)}/2=-\frac{J_0}{2N}[(N-2s_{\mu})^2-N],
\end{align}
where $s_{\mu}=0,\dots,N$ denotes the number of inverted spins in $\boldsymbol{\phi}_{\mu}$.

\begin{figure}[tb]
\centering
\includegraphics[width=.49\textwidth]{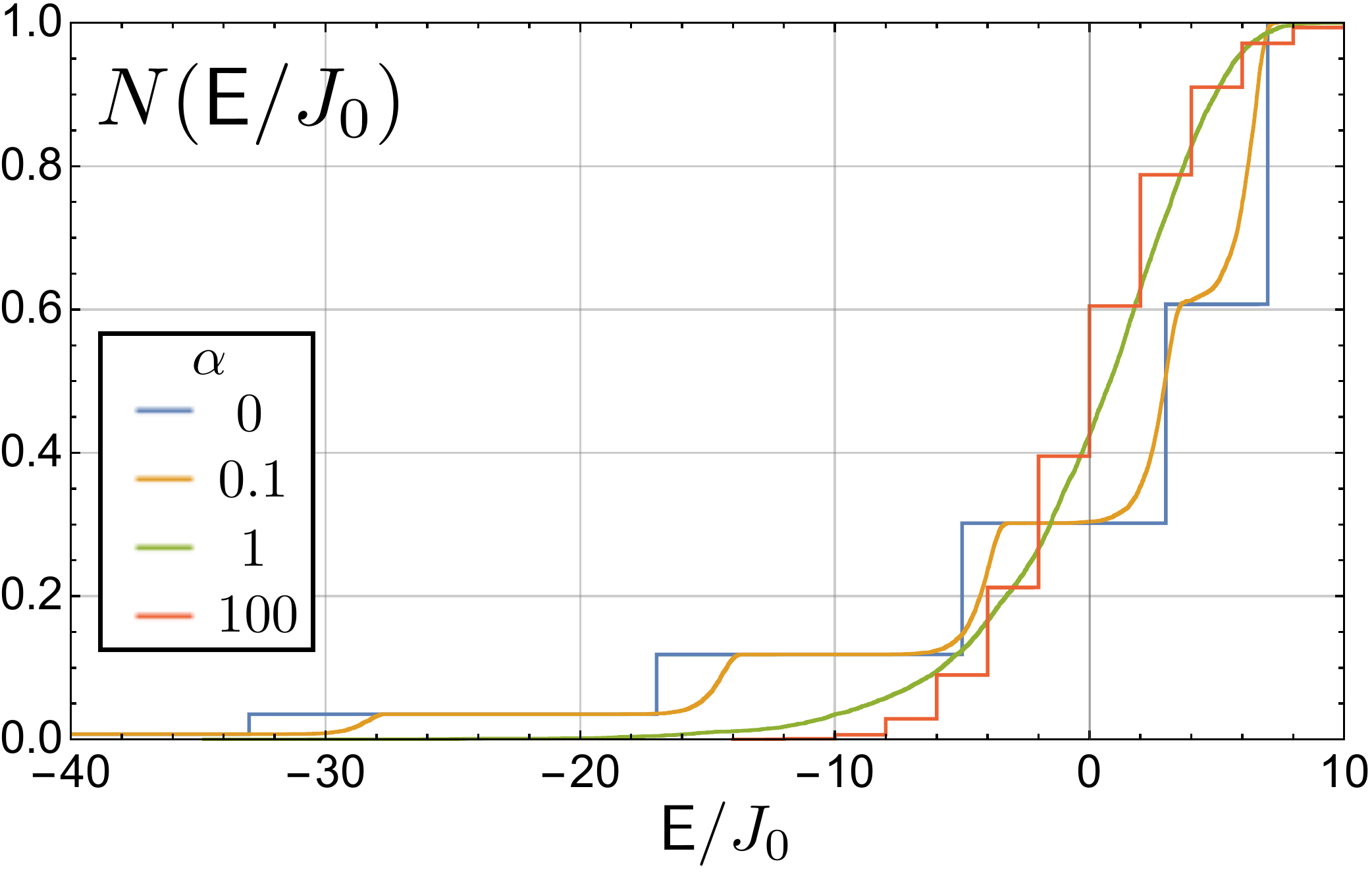}
\caption{The counting function (\ref{eq.norm_counting}) in the ferromagnetic regime ($J_0>0$, $B=0$) for $S=1/2$ reflects a strongly degenerate, quadratically spaced energy spectrum of the Lipkin-Meshkov-Glick model at $\alpha=0$, described by Eq.~(\ref{eq.LMGspectrum}). As the range of the interaction decreases, i.e., as $\alpha$ increases, the spectrum evolves into a broadly distributed energy distribution, especially at values of $\alpha\approx 1$, and finally approaches the equally spaced, and, again, strongly degenerate spectrum of the Ising model at $\alpha=\infty$, described by Eq.~(\ref{eq.Isingspectrum}), which is symmetric around zero. The spectra are obtained using the exact 
semiclassical result~(\ref{eq.expscstsb0}) for $N=15$ spins.}
\label{fig.dos}
\end{figure}

The normalized counting function 
\begin{align}
\label{eq.norm_counting}
N(\mathsf{E})=2^{-N}\sum_{i=0}^{2^N-1}\theta(\mathsf{E}-\mathsf{E}_i)
\end{align}
is obtained from the exact variational ansatz for $B=0$, and is shown in Fig.~\ref{fig.dos} for different values of $\alpha$. We observe that, as a function of $\alpha$, the spectrum at $B=0$ interpolates smoothly between the two strongly degenerate cases of a quadratically spaced sequence of eigenvalues at $\alpha=0$ and a harmonic spectrum, which is symmetric around zero at $\alpha=\infty$. For intermediate values of $\alpha$, the energy levels are broadly distributed between the two extreme values $\mathsf{E}_{\mathrm{e}}=-N\mathsf{J}_{\mathrm{e}}$ and $\mathsf{E}_{\mathrm{a}}=-N\mathsf{J}_{\mathrm{a}}$, where 
\begin{align}
\label{eq:SpecLowBound}
\mathsf{J}_{\mathrm{e}}=\frac{J_0}{N}\sum_{\substack{i,j=1\\ (i<j)}}^N\frac{1}{|i-j|^{\alpha}}
\end{align}
and
\begin{align}
\label{eq:SpecUpBound}
\mathsf{J}_{\mathrm{a}}=\frac{J_0}{N}\sum_{\substack{i,j=1\\ (i<j)}}^N\frac{(-1)^{i+j}}{|i-j|^{\alpha}}
\end{align}
are generated from Eq.~\eqref{eq.jeff} by means of an equal mean-field configuration of parallel spins, and an alternating mean-field configuration where the spin orientation of neighboring spins is inverted, respectively. These two configurations are depicted in Fig.~\ref{fig.ConfigSpin}. The latter describes a two-fold degenerate ground state configuration of an anti-ferromagnet as well as the highest excited states of a ferromagnet. 

\begin{figure}[tb]
\centering
\includegraphics[width=.49\textwidth]{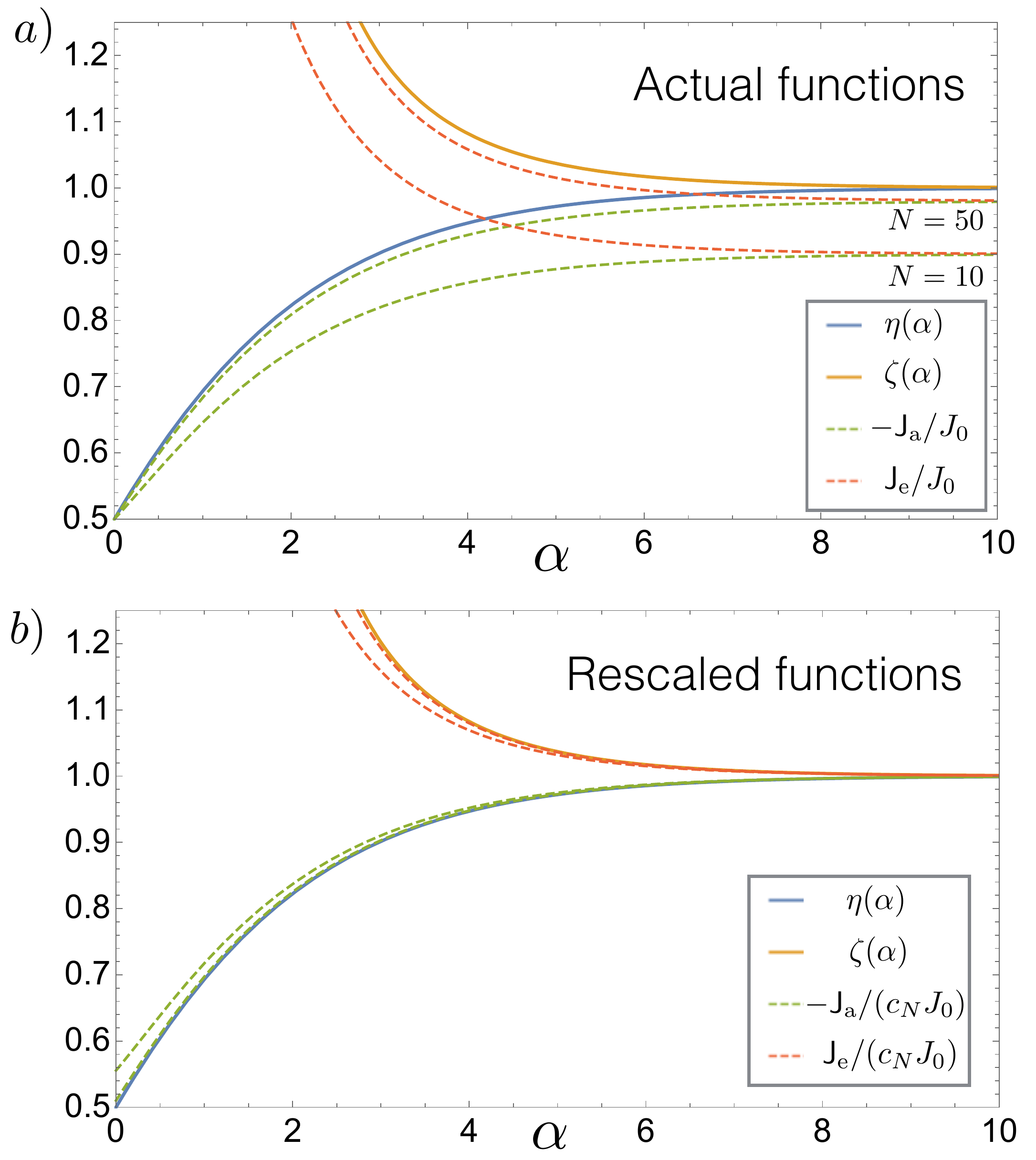}
\caption{a) In the thermodynamic limit ($N\rightarrow\infty$), the effective spin-spin coupling $\mathsf{J}_\mathrm{a}$ [$\mathsf{J}_\mathrm{e}$] of the ground state configuration is given by $\eta(\alpha)$ [$\zeta(\alpha)$] if $J_0<0$ [$J_0>0$], which coincide for $\alpha\rightarrow\infty$. b) Finite-size effects (see dashed lines for $N=10$ and $N=50$) can be compensated with the correction term $c_N=1-1/N$. The rescaled effective spin-spin couplings $\mathsf{J}_{\mathrm{a}}/(J_0c_N)$ and $\mathsf{J}_{\mathrm{e}}/(J_0c_N)$ collapse onto the thermodynamic limit, except for deviations at very small values of $\alpha$.}
\label{fig.jeffuniversal}
\end{figure}

Let us briefly discuss the behavior of $\mathsf{J}_{\mathrm{e}}$ and $\mathsf{J}_{\mathrm{a}}$ in the 
thermodynamic limit $N\rightarrow\infty$. We rewrite
\begin{align}\label{eq.jeconv}
\mathsf{J}_{\mathrm{e}}&=\frac{J_0}{N}\sum_{k=1}^N\frac{N-k}{k^{\alpha}}=J_0\sum_{k=1}^N\frac{1}{k^{\alpha}}+\frac{J_0}{N}\sum_{k=1}^N\frac{1}{k^{\alpha-1}}.
\end{align}
For $\alpha>1$, we have $\zeta(\alpha)=\lim_{N\rightarrow\infty}\sum_{k=1}^Nk^{-\alpha}$, where $\zeta(\alpha)$ denotes Riemann's zeta function~\cite{Edwards}. In this case the second term approaches zero, since $\sum_{k=1}^Nk^{1-\alpha}\sim\mathcal{O}(N^{2-\alpha})$ \cite{Cannas}. Hence, for $\alpha>1$, we have
\begin{align}
\label{eq:SpecLowBoundTDL}
\lim_{N\rightarrow\infty}\mathsf{J}_{\mathrm{e}}=J_0\zeta(\alpha),
\end{align}
whereas for $\alpha\leq 1$ the sum diverges and the spectrum becomes unbounded. At $\alpha=1$ the divergence is logarithmic in $N$. 

Employing an analogous rearrangement of terms, we find for {$\alpha> 0$},
\begin{align}
\label{eq:SpecUpBoundTDL}
\lim_{N\rightarrow\infty}\mathsf{J}_{\mathrm{a}}=-J_0\eta(\alpha),
\end{align}
where $\eta(\alpha)=\lim_{N\rightarrow\infty}\sum_{k=1}^N(-1)^kk^{-\alpha}$ is Dirichlet's eta function ($\alpha>0$)~\cite{Apostol}, which is related to Riemann's zeta function by $\eta(\alpha)=(1-2^{1-\alpha})\zeta(\alpha)$. 
Evaluating the sum explicitly at $\alpha=0$ yields $\lim_{N\rightarrow\infty}\mathsf{J}_{\mathrm{a}}=-J_0/2$ which coincides with 
the analytic continuation~\cite{Cartan}, $\eta(0)=1/2$, allowing us to extend the above equality to all $\alpha\geq 0$. 

For $\alpha=\infty$ both $\mathsf{J}_{\mathrm{a}}$ and $\mathsf{J}_{\mathrm{e}}$ coincide in magnitude in the thermodynamic limit, since $\lim_{\alpha\rightarrow\infty}\zeta(\alpha)=\lim_{\alpha\rightarrow\infty}\eta(\alpha)=1$, and, thus, the Ising spectrum is bounded between $\pm NJ_0$, which, in the considered limit of large $N$, is consistent with Eq.~(\ref{eq.Isingspectrum}). The convergence towards the thermodynamic limit is displayed in Fig.~\ref{fig.jeffuniversal}. To leading order, finite-size effects are caused by the $1/N$ prefactor of the second term of Eq.~(\ref{eq.jeconv}), and of the corresponding alternating expression for $\mathsf{J}_{\mathrm{a}}$. Hence, to a good approximation, these finite-size effects are compensated by a factor $c_N=1-1/N$, and small deviations can only be observed when $\alpha$ is very small, as is shown in Fig.~\ref{fig.jeffuniversal} b).

\subsection{Paramagnetic spectrum at $J_0=0$}\label{sec.para}
After discussing the \mbox{(anti-)}ferromagnetic spectrum, we turn to the opposite limit of very strong external magnetic fields, by setting the spin-spin coupling to zero: $J_0=0$. The Hamiltonian
\begin{align}
H_{\mathrm{ex}}=\left. H\right|_{J_0=0}= - 2B \sum_{i=1}^N S_y^{(i)},
\end{align}
describing the interaction with the external field, is independent of $\alpha$. The spectrum of $H_{\mathrm{ex}}$ is easily found, e.g., by employing a treatment in complete analogy to the one shown in the beginning of the preceding Section: We introduce product states $|S,\mathbf{m}_y\rangle$ of local eigenstates $|S,m^{(i)}_y\rangle$ of $S_y^{(i)}$, characterized by a vector $\mathbf{m}_y=(m_y^{(1)},\dots,m_y^{(N)})$ of eigenvalues $m^{(i)}_y=-S,\dots,S$. This leads to the eigenvalue equation
\begin{align}\label{eq.exactparaspectrum}
H_{\mathrm{ex}}|S,\mathbf{m}_y\rangle=- 2B \sum_{i=1}^N m_y^{(i)}|S,\mathbf{m}_y\rangle.
\end{align}
The resulting spectrum is harmonic and elementary excitations are given by \textit{spin flips} against the magnetic field in $y$-direction. Recalling Eqs.~(\ref{eq.exp1})--(\ref{eq.exp3}), we see that the previously introduced inversion of a spin, $\phi\rightarrow -\phi$, which corresponds to a mirror reflection at the $y$-axis, does not change the expectation value of the paramagnetic energy term. However, the configurations $\boldsymbol{\phi}_{\mu}=(\phi_{\mu}^{(1)},\phi_{\mu}^{(2)},\dots,\phi_{\mu}^{(N)})$ are able to account for such excitations via spin \textit{flips}, defined by the operation $\phi_{\mu}^{(i)}\rightarrow\phi_{\mu}^{(i)}+\pi$, which describes a combined mirror reflection at the $x$ and $y$-axes. We again describe the entire spin chain configuration in terms of a single angle $\phi$, and introduce $\xi_{\mu}^{(i)}=\pm 1$ through $\cos\phi_{\mu}^{(i)}=\xi_{\mu}^{(i)}\cos\phi$ to label the presence or absence of a spin flip at position $i$.

Indeed, employing the spin coherent states~(\ref{eq.trialstate}), combined with the above constraints, generates the following energy expectation values [see Eq.~(\ref{eq.averageenergygeneral})] for $J_0=0$ and at $\phi=0$: 
\begin{align}
E_{\mu}=- \frac{B}{SN} \sum_{i=1}^N l_{\mu}^{(i)}\xi_{\mu}^{(i)},
\end{align}
which, due to $-S\leq l_{\mu}^{(i)}\xi_{\mu}^{(i)}\leq S$, reproduce the full spectrum, as given in Eq.~(\ref{eq.exactparaspectrum}). According to Eqs.~(\ref{eq.exp1})--(\ref{eq.exp3}) the angle $\phi=0$ reflects the polarization of the spins along the $y$-direction of the external field.

Let us focus again on the special case of $S=1/2$. We then can express the energy eigenvalues for $J_0=0$ as $\mathsf{E}_{\mu}=-N\mathsf{B}_{\mu}$, where the effective magnetic fields are given by
\begin{align}
\mathsf{B}_{\mu}=(N-2k_{\mu})B/N,
\end{align}
and $k_{\mu}=0,\dots,N$ counts the number of flipped spins in the configuration characterized by $\boldsymbol{\phi}_{\mu}$. We recover the well-known equidistant energy levels of a paramagnetic chain. This resembles the spectrum at $B=0$ when $\alpha=\infty$. The difference between the two cases is that for $J_0=0$, there can be between $0$ and $N$ inverted spins, which leads to $N+1$ energy bands, whereas for $B=0$ and $\alpha=0$ there are between $0$ and $N-1$ domain walls, and, thus, only $N$ energy bands. 

Notice that, due to the symmetry of the paramagnetic spectrum with respect to energy zero, a global change of the signs of all energy eigenvalues for arbitrary $J_0$ and $B$ always 
produces the spectrum of the chain for parameter values $-J_0$ and $B$, independently of $\alpha$ and $S$.

\subsection{Semiclassical spectra from analytically determined extrema of one-dimensional energy landscapes: Multi-configurational mean-field approach}\label{sec.mcmfa}
So far, we formulated a variational ansatz in terms of spin coherent states to reproduce the exact spectra when either $B=0$ or $J_0=0$, for arbitrary $\alpha$ and $S$. Starting from a uniformly distributed spin arrangement $(\phi,\dots,\phi)$, we employed combinations of spin inversions $(\phi_{\mu}^{(i)}\rightarrow-\phi_{\mu}^{(i)})$ and spin flips $(\phi_{\mu}^{(i)}\rightarrow \phi_{\mu}^{(i)}+\pi)$ to design excited-state configurations $\boldsymbol{\phi}_{\mu}=(\phi_{\mu}^{(1)},\phi_{\mu}^{(2)},\dots,\phi_{\mu}^{(N)})$. Note that spin flips, $\phi_{\mu}^{(i)}\rightarrow\phi_{\mu}^{(i)}+\pi$, also change the \mbox{(anti-)}ferromagnetic energy expectation value, which can be compensated by an additional inversion, $\phi_{\mu}^{(i)}\rightarrow-\phi_{\mu}^{(i)}+\pi$. In total, each spin can assume one of four different orientations, i.e., $\phi_{\mu}^{(i)}=\phi$, $\phi_{\mu}^{(i)}=-\phi$ (inverted), $\phi_{\mu}^{(i)}= \phi+\pi$ (flipped), and $\phi_{\mu}^{(i)}= -\phi+\pi$ (inverted and flipped). 
Independently of the orientation, each of the spin coherent states also has a tunable length $l_{\mu}^{(i)}$ that can assume discrete values between the minimum value $0$ (if $S$ is integer) or $1/2$ (if $S$ is half-integer) and the maximum value $S=M/2$.

The orientation of each individual spin $\phi_{\mu}^{(i)}$ is parametrized by a single angle $\phi$, and, thus, the variational ansatz can be understood as 
a mean-field approach. Based on the above recipe, we obtain an entire family of mean-field descriptions (\textit{multi-configurational mean-field}), labelled by the index $\mu$, which represents a particular spin configuration. A spin configuration $(\mathbf{l}_{\mu},\boldsymbol{\phi_{\mu}})$, is fully characterized by the two vectors $\mathbf{l}_{\mu}$ and $\boldsymbol{\phi_{\mu}}$, which determine the local lengths and orientations of the spins, respectively. Suitable design of these configurations leads, via Eq.~(\ref{eq.trialstate}), to a series of single-parameter trial states that yield any arbitrary eigenvalue at $B=0$ and, independently, any arbitrary eigenvalue at $J_0=0$. These two eigenvalue solutions will then be attained at different values of the parameter $\phi$: As we saw in the previous Sections, the spin orientation in the paramagnetic phase is given by $\phi=0$, while in the \mbox{(anti-)}ferromagnetic phase we have $\phi=\pm\pi/2$.

\begin{figure}[tb]
\centering
\includegraphics[width=.35\textwidth]{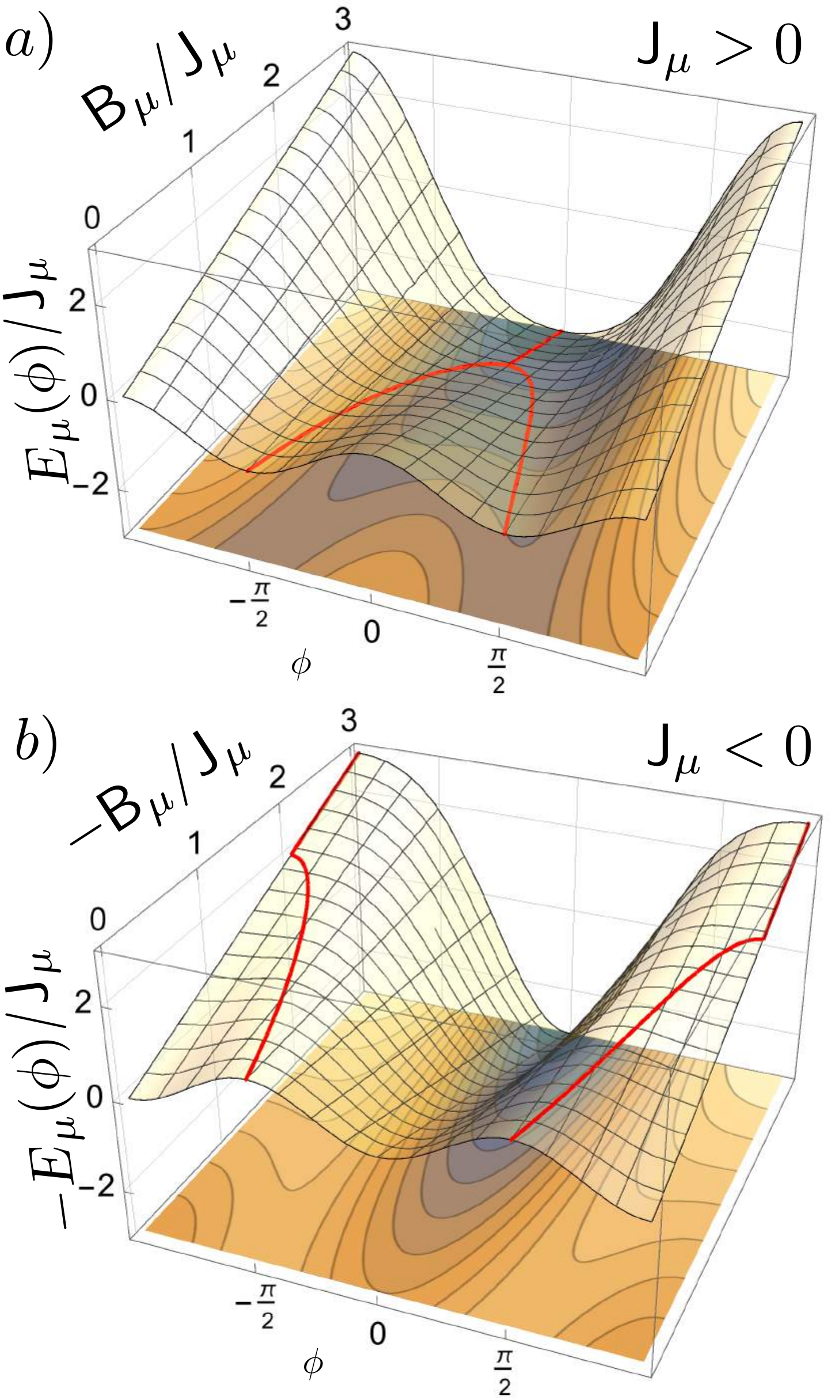}
\caption{a) Energy landscape for positive $\mathsf{J}_{\mu}$ as a function of $\phi$ and $\mathsf{B}_{\mu}$. The energy minimum (red line) represents a semiclassical energy level as a function of $\mathsf{B}_{\mu}\propto B$ [see Eq.~(\ref{eq:bmu})]. The change of the dependence of the minimum energy on $B$ from quadratic to linear is a consequence of the underlying bifurcation into two degenerate but distinct solutions at weak magnetic fields. In the case of the ground state, such a bifurcation represents the semiclassical analogue of the tipping point between the symmetric paramagnetic state, where the symmetry of the Hamiltonian is dictated by the $B$-dependent term, and the symmetry-broken \mbox{(anti-)}ferromagnetic state, where the symmetry of the Hamiltonian is dictated by the $J_0$-term. b) When $\mathsf{J}_{\mu}$ is negative, the energy maximum describes a semiclassical 
energy level and exhibits analogous behavior.}
\label{fig.energymanifold}
\end{figure}

To continuously parametrize the energy spectrum for arbitrary $B$ and $J_0$, we employ the configurations $\boldsymbol{\phi}_{\mu}$ derived above to analyze the semiclassical energy landscape $E_{\mu}(\mathbf{l}_{\mu},\boldsymbol{\phi}_{\mu})$, introduced in Eq.~(\ref{eq.averageenergygeneral}), as a function of $B$ and of the angle $\phi$. For each $\mu$, we obtain a one-dimensional semiclassical energy landscape
\begin{align}\label{eq.energylandscape}
E_{\mu}(\phi)=-\mathsf{J}_{\mu} \sin^2\phi - \mathsf{B}_{\mu} \cos\phi,
\end{align}
with
\begin{align}
\mathsf{J}_{\mu}=\frac{J_0}{S^2N}\sum_{\substack{i,j=1\\(i < j)}}^N \frac{l_{\mu}^{(i)}\epsilon^{(i)}_{\mu}l_{\mu}^{(j)}\epsilon^{(j)}_{\mu}}{|i-j|^{\alpha}},
\end{align}
and
\begin{align}
\label{eq:bmu}
\mathsf{B}_{\mu}=\frac{B}{S N} \sum_{i=1}^N l_{\mu}^{(i)}\xi_{\mu}^{(i)}.
\end{align}

The semiclassical energy landscape~\eqref{eq.energylandscape} is characterized by the effective magnetic field $\mathsf{B}_{\mu}$, which is proportional to $B$, and the effective spin-spin coupling constant $\mathsf{J}_{\mu}$, proportional to $J_0$. These two effective parameters determine the exact spectra in the extreme cases considered before. The resulting energy landscape is depicted as a function of $B$ in Fig.~\ref{fig.energymanifold}. The position of its extremal values shifts as a function of $B$.

For $\mathsf{J}_{\mu}>0$ and $\mathsf{B}_{\mu}>0$, we find the minimal energy
\begin{align}\label{eq.generaljeffmin}
E^{\mathrm{min}}_{\mu}(B)=\begin{cases}-\mathsf{J}_{\mu}-\mathsf{B}_{\mu}^2/(4\mathsf{J}_{\mu}),& \mathsf{B}_{\mu}< 2\mathsf{J}_{\mu}\\-\mathsf{B}_{\mu},& \mathsf{B}_{\mu}\geq2\mathsf{J}_{\mu}\end{cases}.
\end{align}

\begin{figure*}[tbp]
\centering
\includegraphics[width=.95\textwidth]{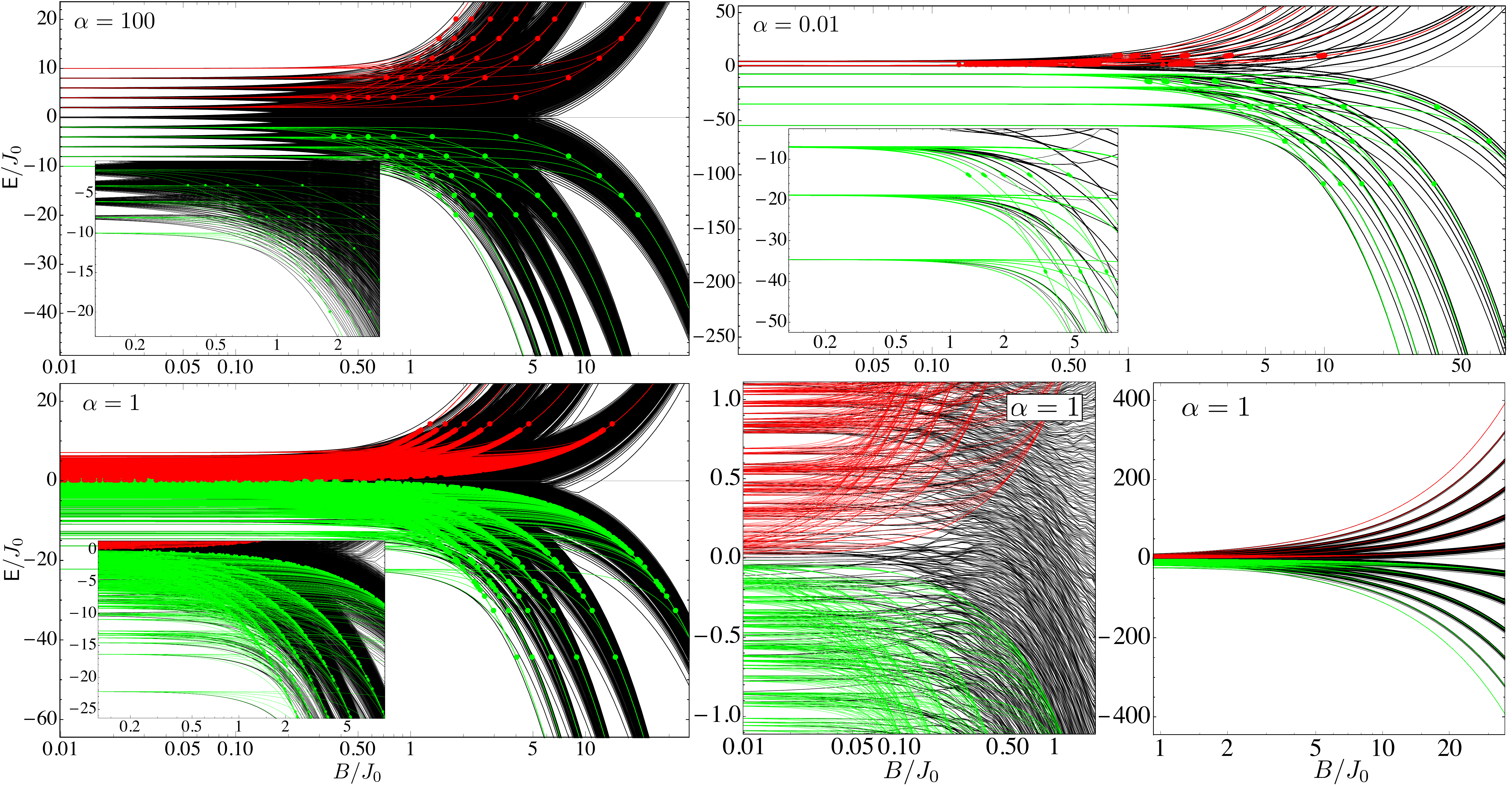}
\caption{Comparison between the exact quantum spectrum (black lines) and the semiclassical energy values, depicted as green and red lines, depending on whether the latter are given by a minimum (see Fig.~\ref{fig.energymanifold}a) or a maximum (Fig.~\ref{fig.energymanifold}b) of the semiclassical energy landscape, respectively. The semiclassical energy levels yield exact results at $B/|J_0|=0$ and $B/|J_0|=\infty$. The dots indicate the points at which the corresponding extremum of the semiclassical energy landscape exhibits a bifurcation, together with a change of its magnetic field dependence from quadratic to linear. Parameters in (\ref{eq.spinchainS}) are $N=11$, $S=1/2$, and $J_0>0$, for different interaction ranges $\alpha$ as indicated. Changing signs on all energy levels yields the corresponding spectra for $-J_0$. The insets zoom into the parameter region where levels reorganize to mediate the transition from the ferro- to the paramagnetic phase.}
\label{fig.semiclassics3}
\end{figure*}

For $\mathsf{B}_{\mu}\geq 2\mathsf{J}_{\mu}$ a unique minimum is identified at $\phi_0=0$, which coincides with the result for the paramagnetic system ($J_0=0$). For $\mathsf{B}_{\mu}< 2\mathsf{J}_{\mu}$ the minimum is two-fold degenerate at $\phi_{\pm}=\pm\arccos \mathsf{B}_{\mu}/(2\mathsf{J}_{\mu})$, which reflects our previous result for the \mbox{(anti-)}ferromagnetic system ($B=0$; see Sec.~\ref{eq.anitferro}), where the two degenerate energy eigenvalues were found at $\phi=\pm\pi/2$. As we continuously scan $B$ [remember, according to (\ref{eq:bmu}), that $\mathsf{B}_{\mu}\propto B$], the position of the minimum changes and a bifurcation occurs, see Fig.~\ref{fig.energymanifold} a). Simultaneously, the minimum energy makes a change from quadratic to linear dependence on $B$, which in turn induces a jump of the energy's second derivative at the bifurcation point $\mathsf{B}_{\mu}= 2\mathsf{J}_{\mu}$.

In the particular case of the semiclassical description of the ground state, the symmetry of the paramagnetic phase is represented by the unique minimum of the energy landscape, whereas two minima express the symmetry-broken character of the \mbox{(anti-)}ferromagnetic phase, in which only one of the two non-symmetric configurations that minimize the semiclassical energy landscape can be realized. In an associated quantum picture, this corresponds 
to two degenerate eigenstates which span the ground-state manifold and lead to quantum superpositions of the two semiclassical configurations.

For $\mathsf{J}_{\mu}<0$ the minimum energy does not depend on $\mathsf{J}_{\mu}$ which would result in an unphysical prediction for the energy at $B=0$. However, in that case the maximum,
\begin{align}\label{eq.generaljeffmax}
E^{\mathrm{max}}_{\mu}(B)=\begin{cases}-\mathsf{J}_{\mu}-\mathsf{B}_{\mu}^2/(4\mathsf{J}_{\mu}),& \mathsf{B}_{\mu}< -2\mathsf{J}_{\mu}\\\mathsf{B}_{\mu},& \mathsf{B}_{\mu}\geq-2\mathsf{J}_{\mu}\end{cases},
\end{align}
represents a physically relevant energy level and shows the same type of bifurcation as the minimum before, as shown in Fig.~\ref{fig.energymanifold} b). The symmetry of the paramagnetic spectrum around zero allows us to restrict to positive values of $\mathsf{B}_{\mu}$. The negative (positive) part of the paramagnetic spectrum is then obtained from $E^{\mathrm{min}}_{\mu}$ $(E^{\mathrm{max}}_{\mu})$ \cite{footnote2}.

\begin{figure}[b]
\centering
\includegraphics[width=.48\textwidth]{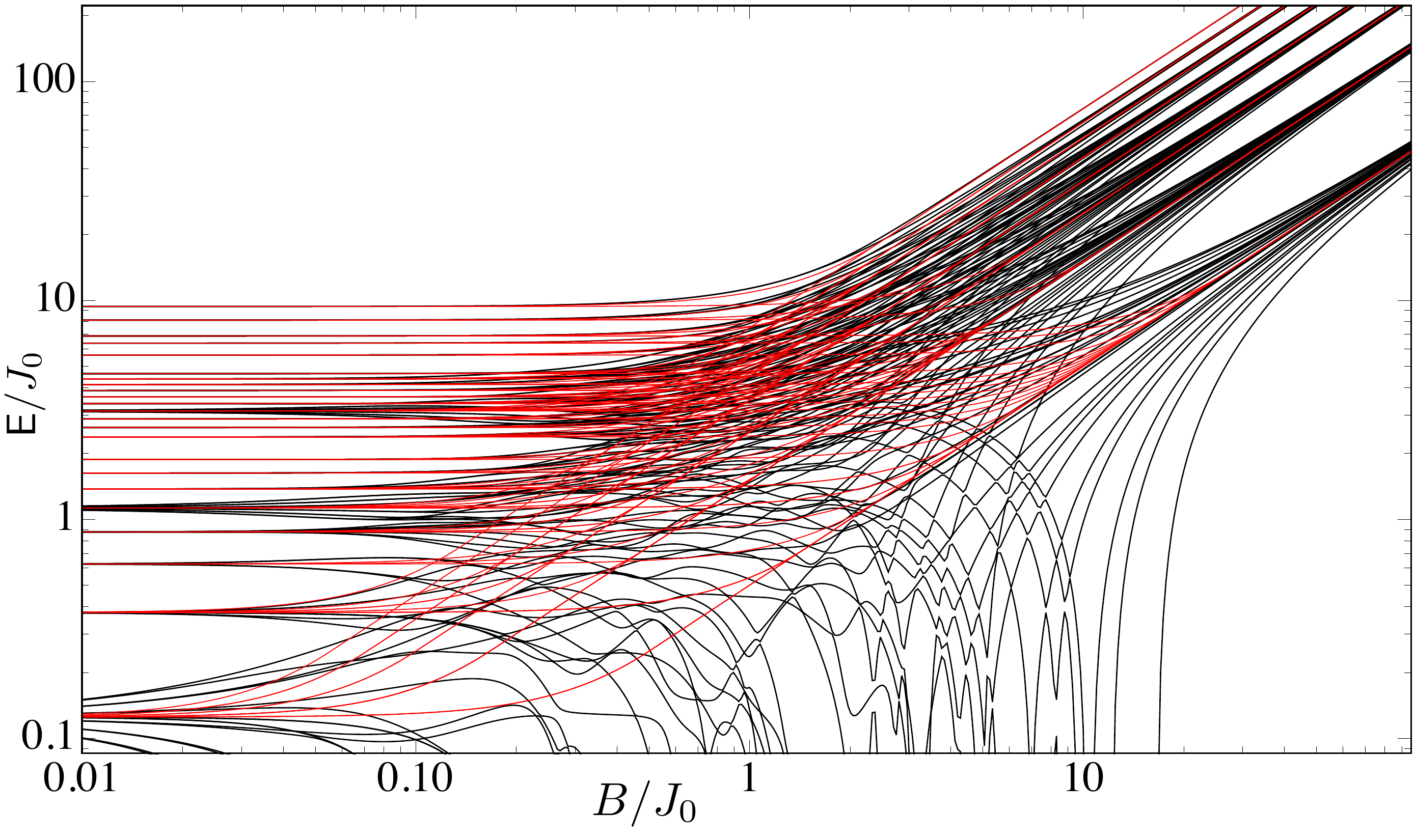}
\caption{Exact quantum spectrum (black lines) and semiclassical energy values (red lines) for the positive energy sector of $H$ with $N=3$, $S=5/2$, $J_0>0$, and $\alpha=1$, as a function of $B/J_0$, on a double-logarithmic scale.}
\label{fig.semiclassicslargeS}
\end{figure}

To summarize, each mean-field configuration that produces a pair of nonzero effective coupling constants $\mathsf{J}_{\mu}$ and $\mathsf{B}_{\mu}$ of equal sign leads to a semiclassical energy landscape, and, depending on the sign of $\mathsf{J}_{\mu}$, its maximum or minimum then characterizes a semiclassical energy level as a function of $B$. The energy levels obtained from Eqs.~(\ref{eq.generaljeffmin}) and (\ref{eq.generaljeffmax}) connect the two exact spectra from $B=0$ to $B\gg |J_0|$. Figures~\ref{fig.semiclassics3} and~\ref{fig.semiclassicslargeS} compare the semiclassical levels to the exact quantum spectrum for the cases $S=1/2$ and $S=5/2$, respectively. Those semiclassical levels obtained from maxima are plotted as red lines, and green lines correspond to semiclassical minima. In the intermediate range, when $B\sim |J_0|$, the quantum spectrum is characterized by an abundance of avoided crossings -- especially when additionally $\alpha\sim 1$ -- which express the incompatibility of the symmetries imposed by the kinetic and by the magnetic term in (\ref{eq.spinchainS}), respectively. This intricate spectral structure reflects the global reorganization of the system eigenstates, within a finite interval of the control parameter, while passing through the phase transition. Such chaotic parametric level dynamics cannot be captured by our semiclassical analysis, since the effective energy manifolds of Fig.~\ref{fig.energymanifold} it is building on are destroyed in this parameter regime.

\subsection{Deviation from the ground state energy for $S=1/2$}
\begin{figure*}[tb]
\centering
\includegraphics[width=.8\textwidth]{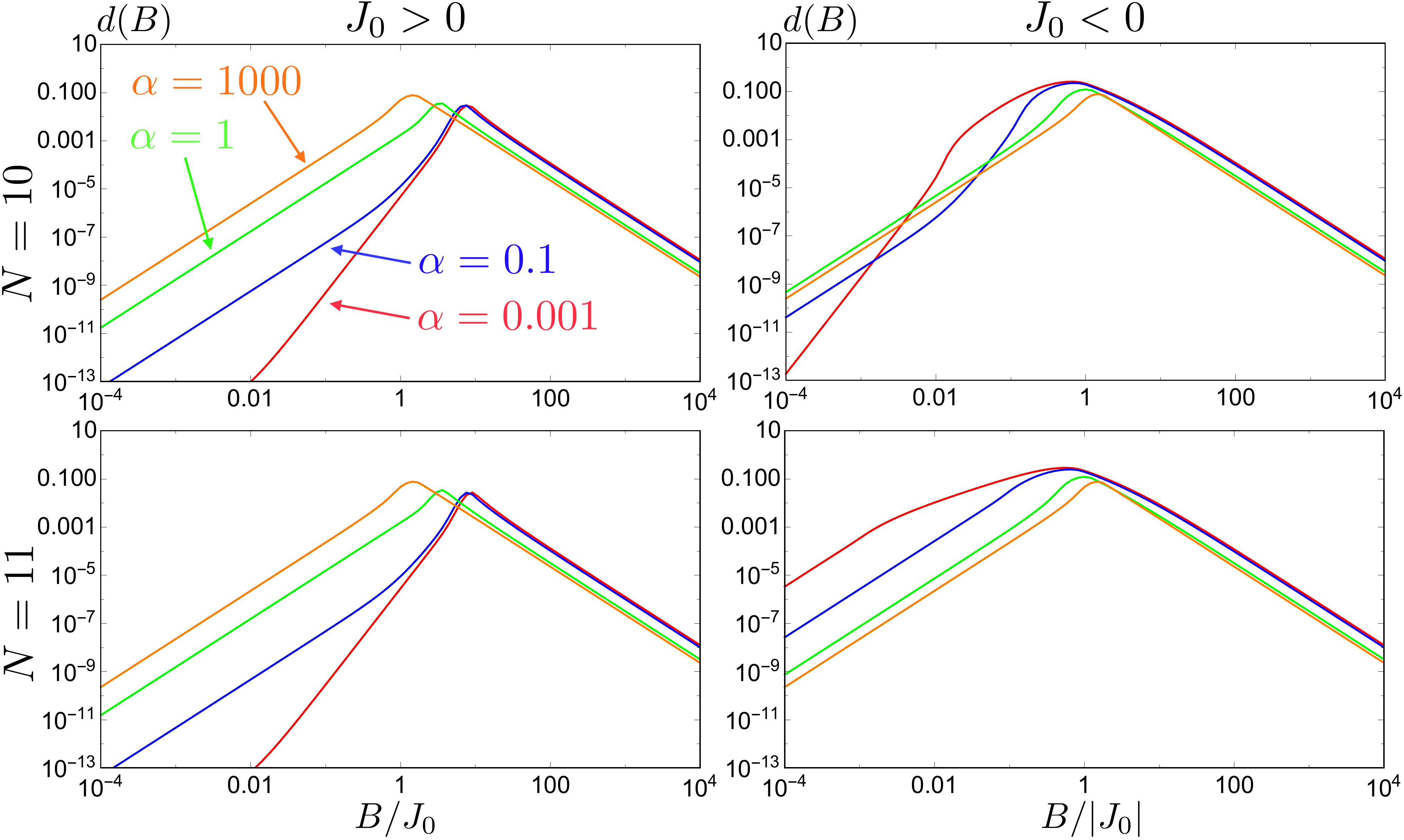}
\caption{Relative deviation $d(B)$, Eq.~(\ref{eq.rel_err}), between the semiclassical and the numerically exact ground state energy for $S=1/2$ and different interaction ranges. In most cases, the exact solution at small and large $B$ is approached as $(B/|J_0|)^{\pm 2}$. Exceptions are found for very long range interactions, which 
converge faster towards the \mbox{(anti-)}ferromagnetic solution, unless $J_0<0$ and $N$ odd. The strongest deviation is found in the vicinity of the 
critical point.}
\label{fig.groundstatedev}
\end{figure*}

We quantify the deviation between the semiclassical results and the exact quantum spectrum through the relative deviation of the respective ground-state energies. We introduce
\begin{align}
\label{eq.rel_err}
d(B)=\frac{E^{\mathrm{min}}(B)-E_0(B)}{|E_0(B)|},
\end{align}
where $E^{\mathrm{min}}(B)$ and $E_0(B)$ denote semiclassical and numerically exact ground state energies, respectively. This quantity is always positive since, when applied to the ground state, our ansatz can be considered as an instance of the Ritz variation principle \cite{Ritz}. We plot $d(B)$ for different parameters in Fig.~\ref{fig.groundstatedev}, on a double-logarithmic scale. The semiclassical ground-state energy approaches the exact paramagnetic result generally as $(B/|J_0|)^{-2}$ for $B\gg |J_0|$, whereas a change of the interaction range $\alpha$ only generates a small constant off-set. Indeed, in this limit, the parameters which determine the spin-spin couplings have vanishing influence. In the opposite limit $B\ll |J_0|$, we also observe a quadratic convergence $\sim(B/|J_0|)^{2}$ towards the exact solution, except for very long range interactions ($\alpha\ll1$), in which case the convergence is even faster for ferromagnets. In the anti-ferromagnetic regime, the convergence rate of long-range interacting systems further depends on whether $N$ is even or odd.

\begin{figure}[tb]
\centering
\includegraphics[width=.49\textwidth]{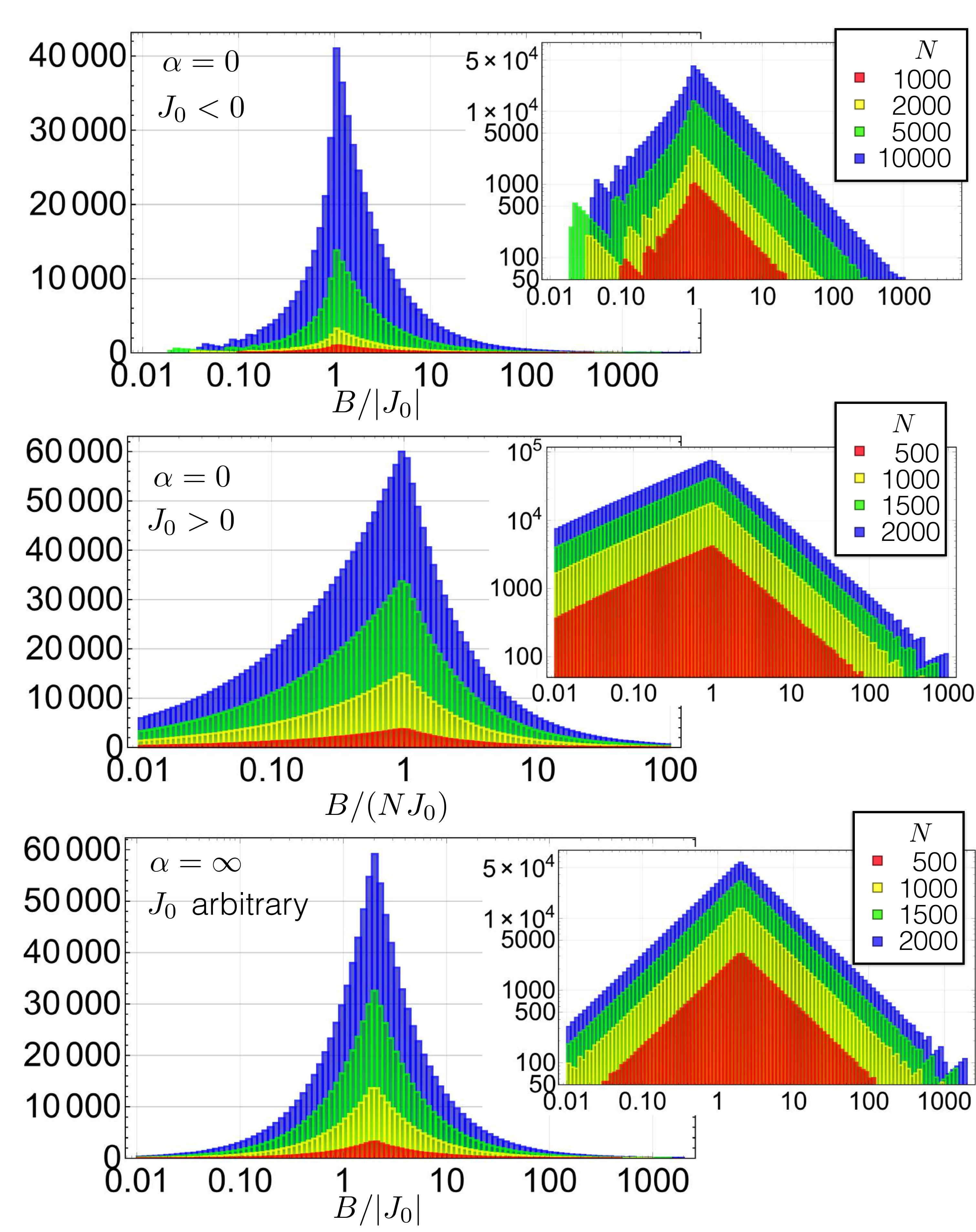}
\caption{(Color online) Logarithmically binned histograms of stable bifurcation points defined by the extremal points of the effective (semiclassical) energy landscapes (see Fig.~\ref{fig.energymanifold}), for the special cases of $\alpha=0$ (Lipkin-Meshkov-Glick) and $\alpha=\infty$ (Ising).}
\label{fig.histograms}
\end{figure}

In the latter case, we recover quadratic convergence with a noticeable off-set compared to interactions of shorter range. The maximum deviation is found at intermediate values of $B\sim |J_0|$, i.e., in the vicinity of the phase transition. Because of the symmetry properties of the Hamiltonian (\ref{eq.spinchainS}) the deviation observed for the ground state at a given value of $J_0$ is equal to that of the highest excited state at $-J_0$.

\subsection{Distribution of excited-state bifurcation points}
The range of intermediate values of $B$ is particularly interesting since this is where we expect the quantum phase transition in the thermodynamic limit, whereas the exact value of the critical point will depend on $\alpha$. Remember that each semiclassical level exhibits a discontinuous second derivative at a specific bifurcation point $B_{\mu}^c$, defined by the condition $\mathsf{B}_{\mu}|_{B=B_{\mu}^c}=2|\mathsf{J}_{\mu}|$ as given by (\ref{eq.generaljeffmin}). In this Section we analyze the distribution of these points as semiclassically determined by the bifurcations of the minima of effective energy landscapes (green dots in Fig.~\ref{fig.semiclassics3}). We will call these stable bifurcation points. As we have noted before, the spectrum for $J_0$ coincides with that for $-J_0$ upon mirroring at energy zero. This implies that the stable bifurcation points of the ferromagnet $(J_0>0)$ coincide with the unstable bifurcation points (those obtained from maxima; red dots in Fig.~\ref{fig.semiclassics3}) of the corresponding anti-ferromagnet $(J_0\rightarrow -J_0)$ and vice versa, which allows us to analyze two systems with one set of $\mathsf{J}_{\mu}$.

The $B_{\mu}^c$ are entirely determined by the effective couplings $\mathsf{J}_{\mu}$ and $\mathsf{B}_{\mu}$, which are straightforwardly obtained. However, a large amount of near-degeneracies in the $\mathsf{J}_{\mu}$, when $\alpha$ is finite but either very large or very small, requires to treat a considerable number of spins ($N\gtrsim 100$), to obtain significant statistics in the intermediate range of $B$-values. For systems of more than $25$ spins this remains computationally challenging, despite the rather simple form of Eq.~(\ref{eq.jeff}). However, the distribution of the $\mathsf{J}_{\mu}$ is easily obtained in the limits $\alpha\rightarrow 0$ and $\alpha\rightarrow\infty$. In these cases, we can resort to the exact results, Eqs.~(\ref{eq.Isingspectrum}) and (\ref{eq.LMGspectrum}), to predict the distribution of bifurcation points analytically, also for large systems. 

Figure~\ref{fig.histograms} displays the distributions of stable bifurcation points on logarithmically binned histograms. In this representation, the histograms show distinct maxima with positions which depend on the sign of $J_0$ and $\alpha$. For the anti-ferromagnetic Lipkin-Meshkov-Glick model ($J_0<0$, $\alpha=0$), we find the most likely bifurcation point at $B=J_0$. In this model, the density of bifurcation points decays as $B^{-1}$ for $B>J_0$, while for $B<J_0$ the distribution does not seem to follow a power-law. For the ferromagnetic case, $J_0>0$, $\alpha=0$, the peak position depends on the system size and is located at $B=NJ_0$. For $B>NJ_0$, we again find a decay of the density as $B^{-1}$, and an increase as $B^{1/2}$ for $B<NJ_0$. Finally, in the Ising limit, $\alpha=\infty$, due to the symmetry of the spectrum, the bifurcation points coincide for both signs of $J_0$. A clear peak can be identified at $B=2J_0$, which is approached from below and above as $B^{1}$ and $B^{-1}$, respectively.

What does this tell us about the quantum bifurcation and the quantum phase transition? Let us start with the case $\alpha=0$: In rescaled units $\bar{J}_0=J_0/N$, the Lipkin-Meshkov-Glick model shows a quantum bifurcation at $B=\bar{J}_0$ when $J_0>0$, which coincides with the histogram's peak position. For $J_0<0$, a first-order transition is found at zero field when rescaled units are used. The histogram peak at $B=J_0$ also approaches zero as $1/N$ when rescaled units are employed. On the other hand, in the Ising limit, $\alpha=\infty$, the maximum of the distribution of bifurcation points deviates from the ground-state critical field by a factor of two, since the quantum phase transition defined by the non-analyticity of the ground state occurs at $B=|J_0|$ \cite{Sachdev}. 

Discrepancies in the Ising limit are not surprising, since the small number of interaction partners for each spin generally limits the performance of mean-field treatments. The energy landscapes characterize the order of the quantum phase transition and provide a qualitative behavior, but for systems of spin-$1/2$ coupled by short-range interactions, they cannot provide a quantitative prediction of, e.g., the exact critical point. In the case of the Ising model, similar observations were made in a study in the context of adiabatic quantum computations \cite{Schaller}.

To summarize, in the present Section we discussed geometrical features of the semiclassical energy landscape $E(\boldsymbol{\phi})$. The stationary points of a series of one-dimensional sections $E_{\mu}(\boldsymbol{\phi}_{\mu})$ [see Eq.~(\ref{eq.energylandscape})] of $E(\boldsymbol{\phi})$ were analyzed analytically. This enabled us to reproduce features of the excitation spectrum at the mean-field level, with an overall good qualitative agreement that even becomes quantitatively exact in parameter regimes far away from the quantum phase transition. The geometrical aspects of those semiclassical energy landscapes further entail essential features of the second-order quantum phase transition, which -- within the presently employed mean-field ansatz -- is associated with the entire excitation spectrum, rather than only the ground state. In the next section, we show that bifurcation points of the semiclassical energy landscape entail dramatic consequences for the quantum corrections in higher orders of the semiclassical expansion. This leads to the observation of quantum bifurcations in the semiclassical limit.

\section{Spin wave theory for excited states}\label{sec.spinwaves}
In perspective of the semiclassical expansion~(\ref{eq:HamTaylorS}), we have, so far, described the semiclassical, spectral features of the spin chain in terms of the lowest-order contribution of Eq.~\eqref{eq:HamTaylorEnergy} and their generalization in Eq.~\eqref{eq.averageenergygeneral}. Now we turn to an analysis of the quadratic quantum fluctuations, which are described by the Hamiltonian Eq.~\eqref{eq.Hquad}, to study the excited states of the spin chain. Considering periodic boundary conditions, $S_{\beta}^{(N+1)}=S_{\beta}^{(1)}$ for $\beta\in\{x,y,z\}$, i.e., a closed ring configuration rather than an open chain, will allow us to obtain an exact analytical description of the dispersion relations of the elementary excitations.

The Hamiltonian describing a ring of long-range interacting spins is given, for arbitrary $N$, by
\begin{align}\label{eq.HarbN}
H^{\mathrm{p}}&=-\frac{J_0}{S}\sum_{i=1}^NS^{(i)}_x\left(\sum_{r=1}^{\lfloor\frac{N-1}{2}\rfloor}\frac{S_x^{(i+r)}+S_x^{(i-r)}}{r^{\alpha}}+\frac{1+(-1)^N}{2}\frac{S_x^{(i+N/2)}}{\left(\frac{N}{2}\right)^{\alpha}}\right)\notag\\
&\quad-2B\sum_{i=1}^NS^{(i)}_y,
\end{align}
where indices such as $(i+r)$, are to be taken as $\bmod \ N$, i.e., $S_{\beta}^{(i+N)}=S_{\beta}^{(i)}$, and $\lfloor x\rfloor$ describes the largest integer $\leq x$. Henceforth, the superscript $\mathrm{p}$ denotes periodic boundary conditions. We simplify the above expression to
\begin{align}\label{eq.spinHpbc}
H^{\mathrm{p}}&=-\frac{
J_0}{S}\sum_{i=1}^{N}\sum_{r\in I^0_N}\frac{S^{(i)}_xS_x^{(i+r)}}{|r|^{\alpha}}-2B\sum_{i=1}^{N}S^{(i)}_y.
\end{align}
where we define the set
\begin{align}I_N=\{-N/2,\dots,-1,0,1,\dots,N/2-1\}
\end{align}
when $N$ is even, and
\begin{align}
I_N=\{-(N-1)/2,\dots,-1,0,1,\dots,(N-1)/2\}
\end{align}
for $N$ odd, respectively, and the summation over $r$ is carried out over $I^0_N=I_N\backslash\{0\}$. As is shown in appendix~\ref{AppendixB}, an analogous derivation as in Section~\ref{sec.bosons} again allows us to express the rotated Hamiltonian $H^{\mathrm{p}}(\boldsymbol{\phi})=U(\boldsymbol{\phi})H^{\mathrm{p}}U^{\dagger}(\boldsymbol{\phi})$ as in Eq.~(\ref{eq:HamTaylorS}), where the semiclassical energy is now given by
\begin{align}\label{eq.meanfieldpbc}
E^{\mathrm{p}}(\boldsymbol{\phi})=-\frac{J_0}{2N}\sum_{i=1}^{N}\sum_{r\in I^0_N}\frac{\sin\phi^{(i)}\sin\phi^{(i+r)}}{|r|^{\alpha}}-\frac{B}{N}\sum_{i=1}^{N}\cos\phi^{(i)},
\end{align}
and linear and quadratic Hamiltonians, $H^{\mathrm{p}}_{\mathrm{L}}(\boldsymbol{\phi})$ and $H^{\mathrm{p}}_{\mathrm{Q}}(\boldsymbol{\phi})$, are determined by the mean-field energy~(\ref{eq.meanfieldpbc}) through similar expressions as before, see Eqs.~(\ref{eq.HlinA}) and (\ref{eq.HquadA}). We look for stationary points of $E^{\mathrm{p}}(\boldsymbol{\phi})$, which generally lead to the disappearance of the linear Hamiltonian ($H^{\mathrm{p}}_{\mathrm{L}}=0$), and study the quantum fluctuations described by
\begin{align}
 \label{eq.HaBosPBC}
&\quad H^{\mathrm{p}}_{\mathrm{Q}}(\boldsymbol{\phi})\notag\\&=-\frac{J_0}{2}\sum_{i=1}^{N}\sum_{r\in I^0_N}\frac{\cos\phi^{(i)}\cos\phi^{(i+r)}}{|r|^\alpha}(a_{i}^{\dagger}+a_{i})(a_{i+r}^{\dagger}+a_{i+r})
            \notag\\&
         \quad   +2J_0\sum_{i=1}^{N}\sum_{r\in I^0_N}\frac{\sin\phi^{(i)}\sin\phi^{(i+r)}}{|r|^\alpha}a_{i}^{\dagger}a_{i}
            \notag\\&
            \quad+2B \sum_{i=1}^{N} \cos\phi^{(i)}a_{i}^{\dagger}a_{i}.
 \end{align}

These expressions are direct extensions of the results for open boundary conditions, presented in Section~\ref{sec.bosons}, to a chain with periodic boundary conditions. The resulting ring configuration implies that the spin that is furthest away from a given spin is found at the opposite side of the ring, rather than at the furthest end of the linear chain, which reduces the maximal distance between two spins. As a consequence, the energy of a finite-sized chain is different from that of a ring of same length. This difference then vanishes in the thermodynamic limit ($N\rightarrow\infty$).

\subsection{Quantum fluctuations around the ferromagnetic ground state}\label{sec.excitationsfmgst}
In Section~\ref{sec.onedim} we developed effective one-dimensional descriptions of the $N$-dimensional configuration space, spanned by the vectors $\boldsymbol{\phi}=(\phi^{(1)},\phi^{(2)}, \dots,\phi^{(N)})$. We now revisit some of these configurations to study the quantum fluctuations around the mean-field results. Notice, however, that the single-parameter configurations $\boldsymbol{\phi}_{\mu}$ which lead to stationary points of the one-dimensional function $E_{\mu}(\boldsymbol{\phi}_{\mu})$ will not necessarily also be stationary configurations of the high-dimensional energy landscape $E^{\mathrm{p}}(\boldsymbol{\phi})$. The simplest single-parameter configuration which \textit{does} lead to a stationary point of $E^{\mathrm{p}}(\boldsymbol{\phi})$ is the uniform configuration $\boldsymbol{\phi}_{\mathrm{e}}=(\phi,\phi,\dots,\phi)$, which in Section~\ref{sec.onedim} was shown to reproduce the ground state energy of a ferromagnetic chain ($J_{0}>0$). This configuration reproduces extremal points of the high-dimensional energy landscape $E^{\mathrm{p}}(\boldsymbol{\phi})$, introduced in Eq.~(\ref{eq.meanfieldpbc}), if the condition [cf. Eq.~(\ref{eq.pbcfirstder})]
\begin{align}
      \label{eq:UniformHamTaylorFixedPoints}
            \left.N\frac{\partial E^{\mathrm{p}}(\boldsymbol{\phi})}{\partial \phi^{(i)}}\right|_{\boldsymbol{\phi}=\boldsymbol{\phi}_{\mathrm{e}}}=\sin\phi\left(B-2\mathsf{J}^{\mathrm{p}}_{\mathrm{e}}\cos\phi \right)\stackrel{!}{=}0,
\end{align}
is satisfied. We have introduced the effective spin-spin interaction constant for periodic boundary conditions
\begin{align}
      \label{eq:PerBondCondEffCoup}
            \mathsf{J}^{\mathrm{p}}_{\mathrm{e}}=\frac{J_{0}}{2}\sum_{r\in I^0_N}\frac{1}{|r|^{\alpha}},
\end{align}
where, again, the subscript $\mathrm{e}$ indicates the equal mean-field configuration. Note that for $\alpha>1$, $\mathsf{J}^{\mathrm{p}}_{\mathrm{e}}$ 
has a well defined thermodynamic limit, which coincides with the limit of $\mathsf{J}_{\mathrm{e}}$, defined in Eq.~\eqref{eq:SpecLowBound} 
[see also Eqs.~\eqref{eq.jeconv} and~\eqref{eq:SpecLowBoundTDL}],
\begin{align}\label{eq.limitJe}
\lim_{N\rightarrow\infty}\mathsf{J}^{\mathrm{p}}_{\mathrm{e}}=\lim_{N\rightarrow\infty}\mathsf{J}_{\mathrm{e}}= J_{0}\zeta(\alpha).
\end{align}
Analogous to Section~\ref{sec.mcmfa}, we find the minima of $E^{\mathrm{p}}(\boldsymbol{\phi})$ to be given by $\cos\phi^{\mathrm{c}}_{\mathrm{e}}=B/2\mathsf{J}^{\mathrm{p}}_{\mathrm{e}}$ for $B\leq 2\mathsf{J}^{\mathrm{p}}_{\mathrm{e}}$ and by $\phi^{\mathrm{c}}_{\mathrm{e}}= 0$ for $B>2\mathsf{J}^{\mathrm{p}}_{\mathrm{e}}$. Inserting these solutions $\phi=\phi^{\mathrm{c}}_{\mathrm{e}}\in[-\pi/2,\pi/2]$ into Eq.~\eqref{eq.meanfieldpbc}, we obtain the semiclassical energy [see also Eq.~\eqref{eq.generaljeffmin}]
\begin{align}
\label{eq:FerrMinEnergy}
E^{\mathrm{p},\min}_{\mathrm{e}}(B)=
\begin{cases}-\mathsf{J}^{\mathrm{p}}_{\mathrm{e}}-B^2/(4\mathsf{J}^{\mathrm{p}}_{\mathrm{e}}),& B\leq 2\mathsf{J}^{\mathrm{p}}_{\mathrm{e}}\\-B,& B>2\mathsf{J}^{\mathrm{p}}_{\mathrm{e}}
\end{cases}.
\end{align}

To investigate the quantum fluctuations about this semiclassical result, we consider the Hamiltonian~\eqref{eq.HaBosPBC} for the uniform configuration $\boldsymbol{\phi}_{\mathrm{e}}$,
\begin{align}
      \label{eq:HamTaylorQuadGroundFerr}
           &\quad \left.H^{\mathrm{p}}_{\mathrm{Q}}(\boldsymbol{\phi}_{\mathrm{e}})\right|_{\phi=\phi^{\mathrm{c}}_{\mathrm{e}}}\notag\\
          &=-\frac{J_0}{2}\cos^{2}\phi^{\mathrm{c}}_{\mathrm{e}}\sum_{i=1}^{N}\sum_{r\in I^0_N} \frac{1}{|r|^\alpha}(a_{i}^{\dagger}+a_{i}) (a_{i+r}^{\dagger}+a_{i+r})
            \nonumber\\&\quad
            +\left(4\mathsf{J}^{\mathrm{p}}_{\mathrm{e}}\sin^{2}\phi^{\mathrm{c}}_{\mathrm{e}}
            +2B\cos\phi^{\mathrm{c}}_{\mathrm{e}}\right) \sum_{i=1}^{N} a_{i}^{\dagger}a_{i}.
\end{align}
We map this Hamiltonian onto the reciprocal space introducing the Fourier transformed operators
\begin{align}
	a_l=\frac{1}{\sqrt{N}}\sum_{k}A_k e^{i kl}
	\label{eq:DFT}.
\end{align}
The periodic boundary conditions $a_{l}=a_{l+N}$ imply that the quasimomenta are quantized, i.e., $k=2n\pi/N$ where $n\in I_N$. In the thermodynamic limit $N\rightarrow\infty$, the quasimomentum becomes a continuous variable $k\in[-\pi,\pi]$. We now transform the Hamiltonian~\eqref{eq:HamTaylorQuadGroundFerr} into reciprocal space. By using the commutation relations $[A_k,A^{\dagger}_{k'}]=\delta_{kk'}$, we eventually obtain
\begin{align}
      \label{eq:HamTaylorQuadGroundRecipFerr}
&\quad \left.H^{\mathrm{p}}_{\mathrm{Q}}(\boldsymbol{\phi}_{\mathrm{e}})\right|_{\phi=\phi^{\mathrm{c}}_{\mathrm{e}}}\notag\\&=-J_0\cos^{2}\phi^{\mathrm{c}}_{\mathrm{e}}\sum_{k} C_{\alpha}^{(N)}(k)\left(A_{k}^{\dagger}A_{-k}^{\dagger}+A_{k}^{\dagger}A_{k}\right.\notag\\&\hspace{3.7cm}\left.+A_{-k}A^{\dagger}_{-k}+A_{k}A_{-k}\right)
            \nonumber\\&\quad
            +(2\mathsf{J}^{\mathrm{p}}_{\mathrm{e}}\sin^{2}\phi^{\mathrm{c}}_{\mathrm{e}}
            +B\cos\phi^{\mathrm{c}}_{\mathrm{e}})\sum_{k} (A_{k}^{\dagger}A_{k}+A_{-k}^{\dagger}A_{-k}).
\end{align}
We introduced the function $C_{\alpha}^{(N)}(k)=(1/2)\sum_{r\in I^0_N} \cos kr/|r|^\alpha$, which in the thermodynamic limit converges to the Clausen function, $\lim_{N\rightarrow\infty}C_{\alpha}^{(N)}(k)=\text{Re}[\mathrm{Li}_{\alpha}(e^{i k})]$, where $\mathrm{Li}_{s}(z)=\sum_{n=1}^{\infty}z^n/n^s$ is the polylogarithm of order $s$~\cite{AndrewsLerch, AbramowitzStegun}.

In order to diagonalize the Hamiltonian~\eqref{eq:HamTaylorQuadGroundRecipFerr}, we consider the canonical form
\begin{align}
      \label{eq:HamTaylorQuadGroundRecipPosit}
            &\quad \left.H^{\mathrm{p}}_{\mathrm{Q}}(\boldsymbol{\phi}_{\mathrm{e}})\right|_{\phi=\phi^{\mathrm{c}}_{\mathrm{e}}}\notag\\&=\sum_{k}\left[F_{\mathrm{e}}(k)(A_{k}^{\dagger}A_{k}+A_{-k}A_{-k}^{\dagger})+G_{\mathrm{e}}(k)(A_{k}^{\dagger}A_{-k}^{\dagger}+A_{-k}A_{k})\right]\notag\\
            &\quad+NE_{\mathrm{e}}^0,
\end{align}
where we defined the real-valued functions
\begin{align}
G_{\mathrm{e}}(k)&=-J_0\cos^{2}\phi^{\mathrm{c}}_{\mathrm{e}}C_{\alpha}^{(N)}(k),\\
F_{\mathrm{e}}(k)&=G_{\mathrm{e}}(k)-E_{\mathrm{e}}^0,\\
E_{\mathrm{e}}^0&=-2\mathsf{J}^{\mathrm{p}}_{\mathrm{e}}\sin^{2}\phi^{\mathrm{c}}_{\mathrm{e}}-B\cos\phi^{\mathrm{c}}_{\mathrm{e}}.\label{eq.ee0}
\end{align}
Now we can diagonalize the Hamiltonian by means of a Bogoliubov transformation: Introducing bosonic operators $\gamma_{\pm k},\gamma_{\pm k}^{\dagger}$ with
\begin{align}
\begin{pmatrix}A_k\\A^{\dagger}_{-k}\end{pmatrix}=\begin{pmatrix}\cosh\theta_k & \sinh\theta_k\\ \sinh\theta_k & \cosh\theta_k\end{pmatrix}\begin{pmatrix}\gamma_k\\\gamma^{\dagger}_{-k}\end{pmatrix}
\end{align}
and $\tanh 2\theta_k=-G_{\mathrm{e}}(k)/F_{\mathrm{e}}(k)$ leads to
\begin{align}
            \left.H^{\mathrm{p}}_{\mathrm{Q}}(\boldsymbol{\phi}_{\mathrm{e}})\right|_{\phi=\phi^{\mathrm{c}}_{\mathrm{e}}}&=\frac{1}{2}\sum_{k}\varepsilon_{\mathrm{e}}(k)\left(\gamma_{k}^{\dagger}\gamma_{k}+\gamma_{-k}^{\dagger}\gamma_{-k}+1\right)
            +NE_{\mathrm{e}}^0
            \nonumber \\
            &=\sum_{k}\varepsilon_{\mathrm{e}}(k)\left(\gamma_{k}^{\dagger}\gamma_{k}+\frac{1}{2}\right)+NE_{\mathrm{e}}^0,
\end{align}
where $\varepsilon_{\mathrm{e}}(k)=2\sqrt{F_{\mathrm{e}}(k)^2-G_{\mathrm{e}}(k)^2}$. The energy offset $E_{\mathrm{e}}^0$ stems from the second derivative of the mean-field term $E^{\mathrm{p}}(\boldsymbol{\phi})$. Its contribution to the $1/\sqrt{2S}$-expansion of $H^{\mathrm{p}}(\boldsymbol{\phi})$ is on the order of $N$, and therefore small compared to the mean-field energy $E^{\mathrm{p}}(\boldsymbol{\phi})$, which contributes in on the order of $2SN$, see Eq.~(\ref{eq.expansionp}). 

\begin{figure}[tb]
\centering
\includegraphics[width=.45\textwidth]{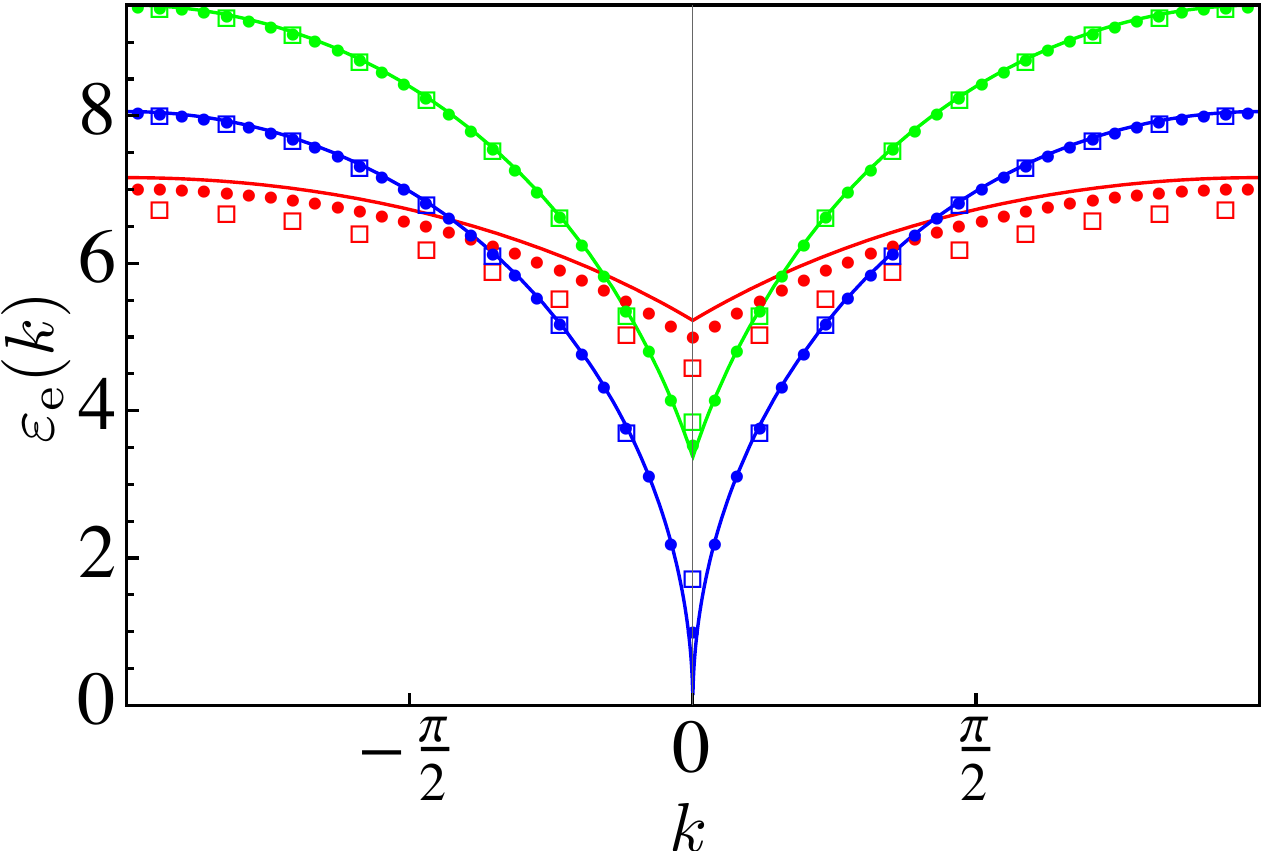}
\caption{Energy dispersion relation $\varepsilon_{\mathrm{e}}(k)$ and finite-size effects for spin-wave quantum corrections to the ferromagnetic ground state. The plot displays the parameters $\alpha=2$ and $B/J_0=2$ (red), $B/J_0=2\zeta(2)$ (blue), and $B/J_0=4$ (green) for $N=17$ (squares), $N=51$ (dots), and $N=\infty$ (lines).}
\label{fig.disprelFinite}
\end{figure}

The properties of the elementary spin-wave excitations \cite{Yosida} are determined by the dispersion relation $\varepsilon_{\mathrm{e}}(k)$, which is depicted in Fig.~\ref{fig.disprelFinite} for finite values of $N$, as well as in the limit $N\rightarrow\infty$. Most interestingly, when the condition $E_{\mathrm{e}}^0=2G_{\mathrm{e}}(k)$ is satisfied, we observe gapless excitations, i.e., $\varepsilon_{\mathrm{e}}(k)=0$. This implies that the spin system allows for the creation of excitations from the ground state at no energy cost, and the consequent instability precisely characterizes the critical point of the quantum bifurcation. 

\begin{figure}[tb]
\centering
\includegraphics[width=.45\textwidth]{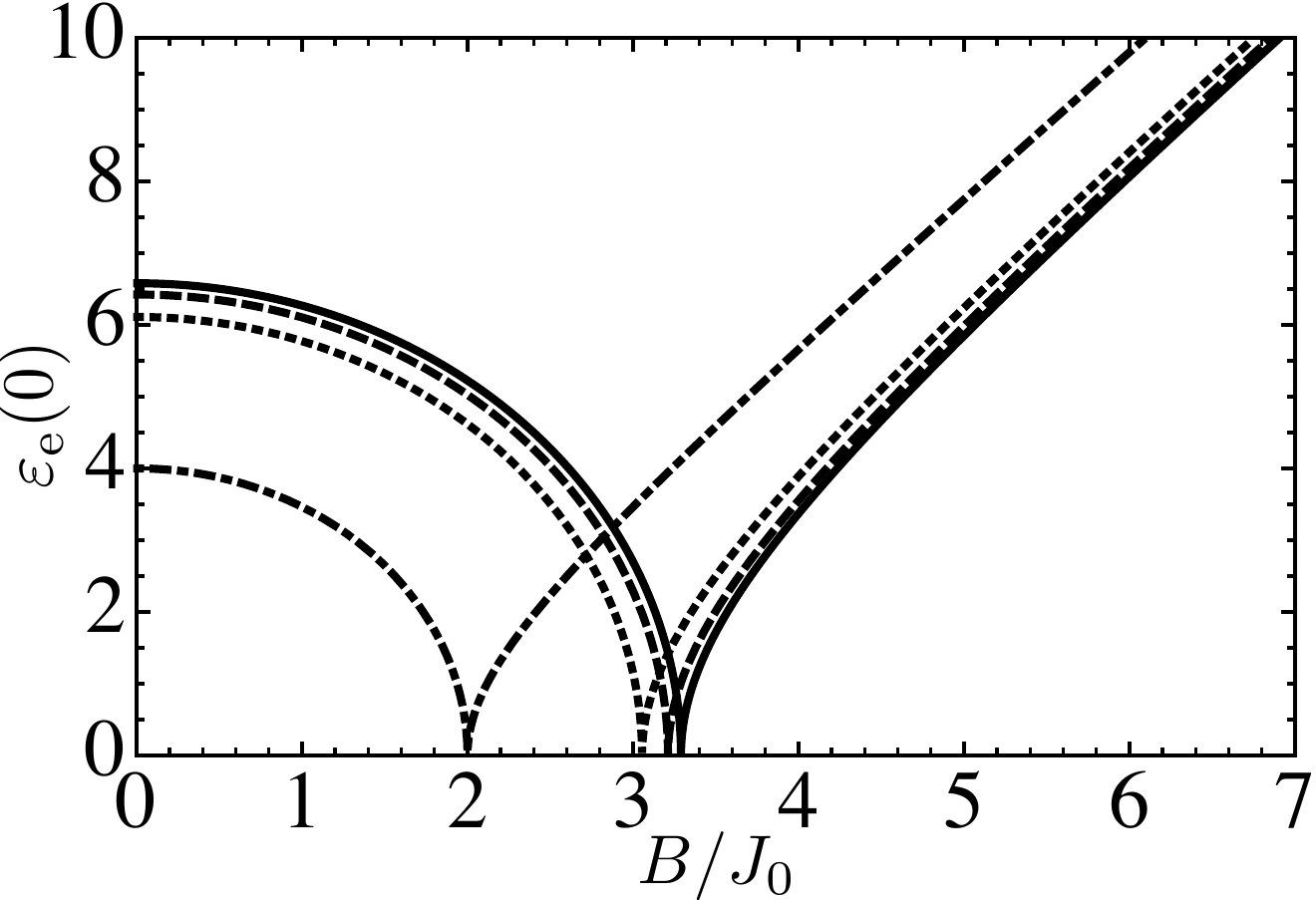}
\caption{Closing of the excitation gap for the ferromagnetic ground state and finite-size effects at $k=0$ for $\alpha=2$ and $N=2$ (dashed-dotted line), $N=17$ (dotted line), $N=51$ (dashed line) and $N=\infty$ (continuous line). The quantum phase transition from ferromagnet to paramagnet occurs when the excitation gap closes at $B^c_{\mathrm{e}}=2\mathsf{J}^{\mathrm{p}}_{\mathrm{e}}$.}
\label{fig.gapFinite}
\end{figure}

The closing of the gap can be observed for the $k=0$ mode when the magnetic field reaches the value $B=B_{\mathrm{e}}^{\mathrm{c}}$, where, hence,
\begin{align}\label{eq.cfieldfm}
B_{\mathrm{e}}^{\mathrm{c}}=2\mathsf{J}^{\mathrm{p}}_{\mathrm{e}}
\end{align}
defines the critical point of the quantum bifurcation. From Eq.~(\ref{eq:FerrMinEnergy}), we observe that this coincides with the bifurcation point of the semiclassical energy landscape, see also Fig.~\ref{fig.energymanifold}. The energy of the $k=0$ mode can be observed in Fig.~\ref{fig.gapFinite} as a function of $B$ for different values of $N$.

This identifies a quantum bifurcation, which, recalling the discussion from Sec.~\ref{sec.QuantumBifurcationbosons}, implies a series of consequences for the elementary excitations close to 
the bifurcation point. In analogy to the diverging correlation length, which is observed in extended lattice systems in the vicinity of a quantum phase transition, we can identify a diverging characteristic length scale as $l_0=1/\sqrt{\varepsilon_{\mathrm{e}}(0)}$. This length scale determines the localization of the resulting $N$-mode Gaussian ground state, which is expected to become strongly squeezed close to the critical point \cite{Brandes03}.

We emphasize here that Eq.~(\ref{eq.cfieldfm}) predicts a sharp quantum bifurcation even without necessarily performing the thermodynamic limit ($N\rightarrow\infty$), since we describe the Hamiltonian in the semiclassical limit $S\gg1$. In fact, the semiclassical limit triggers a sharp discontinuity for all values of $N>1$. This occurs because in the semiclassical limit $S=M/2\gg1$ we consider a large number $M$ of elementary spins at each site $i$, such that the collective spin operators can be mapped to unbounded bosonic degrees of freedom.

\begin{figure}[tb]
\centering
\includegraphics[width=.48\textwidth]{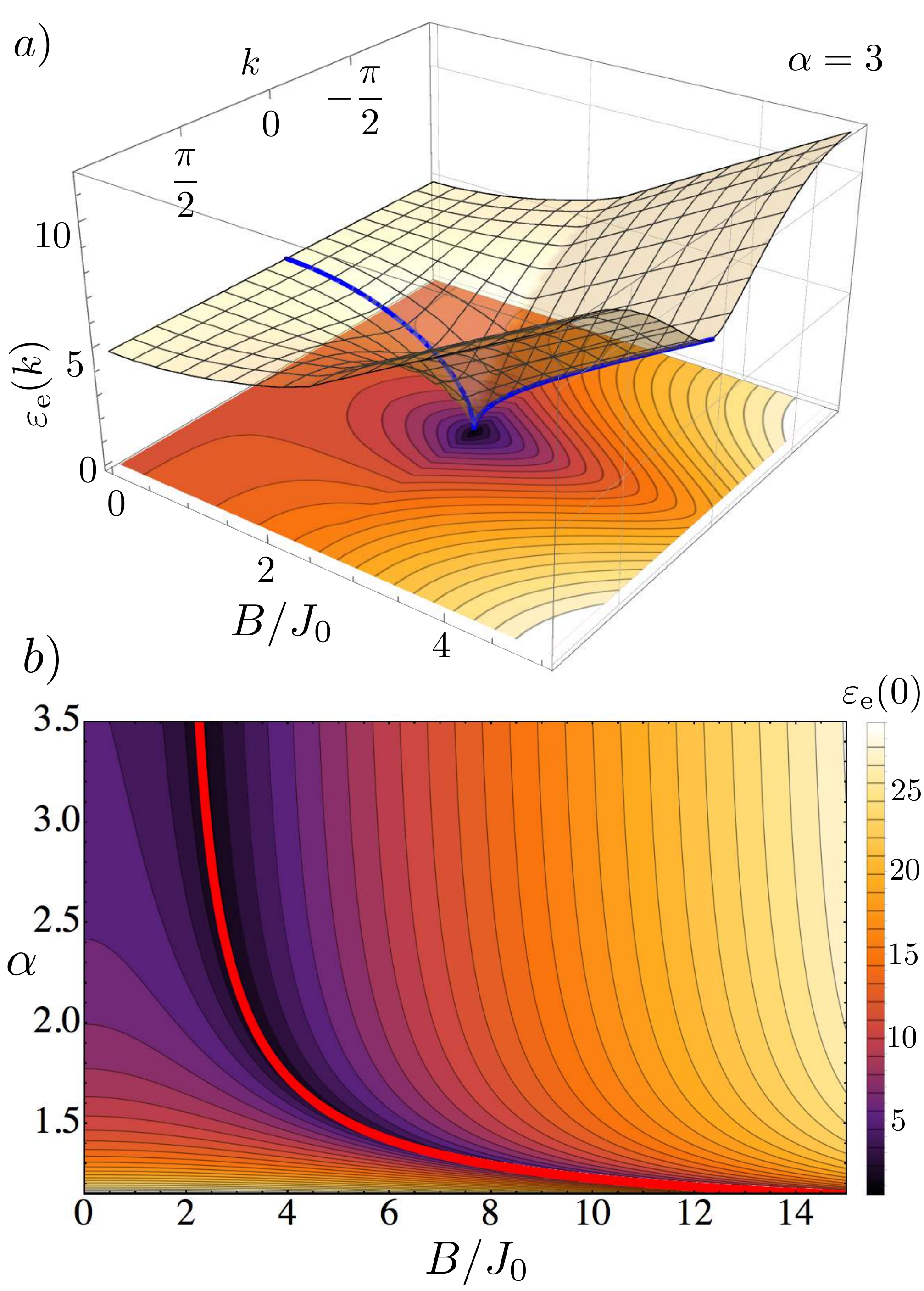}
\caption{a) Energy dispersion relation $\varepsilon_{\mathrm{e}}(k)$ for spin-wave excitations to the ferromagnetic ground state, as a function of $B$ for $\alpha=3$ in the thermodynamic limit $N\rightarrow\infty$. The closing of the excitation gap can be observed at $k=0$ (blue line) at $B=2J_0\zeta(3)\approx 2.4J_0$. b) Energy of the $k=0$ mode as a function of $B$ and $\alpha$. The gap closes at $B=2J_0\zeta(\alpha)$ (red line) for $\alpha>1$.}
\label{fig.energyFerro}
\end{figure}

In the thermodynamic limit $N\rightarrow\infty$, the critical field behaves as $\lim_{N\rightarrow \infty}B_{\mathrm{e}}^{\mathrm{c}}=2J_0\zeta(\alpha)$ when $\alpha>1$, according to Eq.~(\ref{eq.limitJe}).
A complete view of the behavior of the dispersion relation $\varepsilon_{\mathrm{e}}(k)$ as a function of $B$ in the thermodynamic limit is depicted in Fig.~\ref{fig.energyFerro} a), while Fig.~\ref{fig.energyFerro} b) displays the dependence of the critical field on $\alpha$.

\subsection{Quantum fluctuations around the highest excited ferromagnetic state}
The results of the previous Section can be extended to different configurations besides the uniform arrangement of spins $\boldsymbol{\phi}_{\mathrm{e}}$. Next, we consider the alternating configuration $\boldsymbol{\phi}_{\mathrm{a}}$, characterized by $\phi_{\mathrm{a}}^{(i)}=(-1)^{i}\phi$. This configuration was shown to produce the highest excited state of a ferromagnet ($J_0>0$) due to a maximum amount of domain walls, and the ground state of an anti-ferromagnet ($J_0<0$). 

Setting the first derivatives of the semiclassical energy landscape, Eq.~\eqref{eq.pbcfirstder}, to zero, while inserting $\sin\phi^{(i)}=(-1)^{i}\sin\phi$ and $\cos\phi^{(i)}=\cos\phi$, we obtain the condition
\begin{align}
      \label{eq:StaggHamTaylorFixedPoints}
             \left.N\frac{\partial E^{\mathrm{p}}(\boldsymbol{\phi})}{\partial \phi^{(i)}}\right|_{\boldsymbol{\phi}=\boldsymbol{\phi}_{\mathrm{a}}}=(-1)^{i}\sin\phi\left(B-2\mathsf{J}_{\mathrm{a}}^{\mathrm{p}}\cos\phi\right)\stackrel{!}{=}0,
\end{align}
where we defined
\begin{align}
      \label{eq:PerBondCondEffCoupExc}
           \mathsf{J}^{\mathrm{p}}_{\mathrm{a}}=\frac{J_{0}}{2}\sum_{r\in I^0_N}\frac{(-1)^r}{|r|^{\alpha}}.
\end{align}
In the thermodynamic limit we obtain for all $\alpha\geq 0$,
\begin{align}\label{eq.limitJa}
\lim_{N\rightarrow\infty}\mathsf{J}^{\mathrm{p}}_{\mathrm{a}}=\lim_{N\rightarrow\infty}\mathsf{J}_{\mathrm{a}}=-J_{0}\eta(\alpha),
\end{align}
as was shown in Eq.~\eqref{eq:SpecUpBoundTDL}.

As follows immediately from the discussion in Section~\ref{sec.mcmfa}, the solution $\phi^c_{\mathrm{a}}=\pi$ for $B>-2 \mathsf{J}^{\mathrm{p}}_{\mathrm{a}}$ and $\cos\phi^c_{\mathrm{a}}=-B/2 \mathsf{J}^{\mathrm{p}}_{\mathrm{a}}$ for $B\leq -2 \mathsf{J}^{\mathrm{p}}_{\mathrm{a}}$ yields a maximum of $E^{\mathrm{p}}(\boldsymbol{\phi})$ for $J_0>0$, with the value
\begin{align}
\label{eq:AFerrMinEnergy}
E^{p,\max}_{\mathrm{a}}(\phi_{s})=
\begin{cases}- \mathsf{J}^{\mathrm{p}}_{\mathrm{a}}-B^2/(4 \mathsf{J}^{\mathrm{p}}_{\mathrm{a}}),& B\leq -2 \mathsf{J}^{\mathrm{p}}_{\mathrm{a}}\\B,& B>-2 \mathsf{J}^{\mathrm{p}}_{\mathrm{a}}
\end{cases},
\end{align}
in direct correspondence with Eq.~\eqref{eq.generaljeffmax}.

\begin{figure}[tb]
\centering
\includegraphics[width=.48\textwidth]{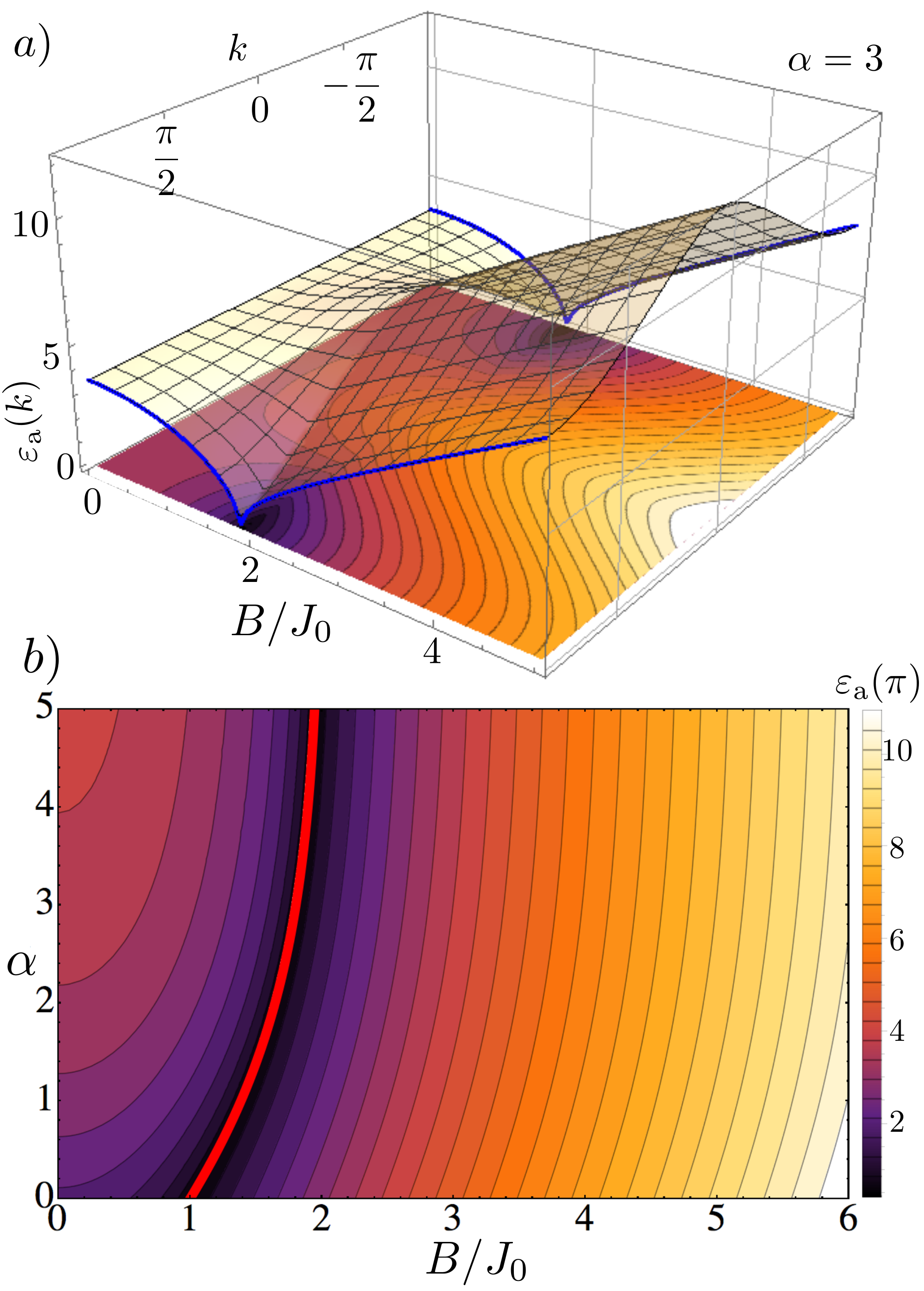}
\caption{a) Energy dispersion relation $\varepsilon_{\mathrm{a}}(k)$ for spin-wave excitations around the highest excited ferromagnetic state, as a function of $B$ for $\alpha=3$ in the thermodynamic limit $N\rightarrow\infty$. 
The closing of the excitation gap now occurs at $k=\pm\pi$ (blue lines). b) Energy of the $k=\pi$ mode as a function of $B$ and $\alpha$. The gap closes at $B=-2J_0\eta(\alpha)$ (red line).}
\label{fig.energyFerroExcited}
\end{figure}

Employing this extremal solution, the quadratic Hamiltonian~\eqref{eq.HaBosPBC} reads
\begin{align}
      \label{eq:HamTaylorQuadStaggered}
           &\quad \left.H^{\mathrm{p}}_{\mathrm{Q}}(\boldsymbol{\phi}_{\mathrm{a}})\right|_{\phi=\phi^{\mathrm{c}}_{\mathrm{a}}}\notag\\
          &=-\frac{J_0}{2}\cos^{2}\phi^{\mathrm{c}}_{\mathrm{a}}\sum_{i=1}^{N}\sum_{r\in I^0_N} \frac{1}{|r|^\alpha}(a_{i}^{\dagger}+a_{i}) (a_{i+r}^{\dagger}+a_{i+r})
            \nonumber\\&\quad
            +\left(4\mathsf{J}^{\mathrm{p}}_{\mathrm{a}}\sin^{2}\phi^{\mathrm{c}}_{\mathrm{a}}
            +2B\cos\phi^{\mathrm{c}}_{\mathrm{a}}\right) \sum_{i=1}^{N} a_{i}^{\dagger}a_{i},
\end{align}
which has exactly the same form as Eq.~\eqref{eq:HamTaylorQuadGroundRecipFerr}. Thus, an analogous derivation yields the following diagonal representation in terms of the Bogoliubov modes $\delta_k$ and $\delta^{\dagger}_k$:
\begin{align}
            \left.H^{\mathrm{p}}_{\mathrm{Q}}(\boldsymbol{\phi}_{\mathrm{a}})\right|_{\phi=\phi^{\mathrm{c}}_{\mathrm{a}}}=\sum_{k}\varepsilon_{\mathrm{a}}(k)\left(\delta_{k}^{\dagger}\delta_{k}+\frac{1}{2}\right)+NE_{\mathrm{a}}^0,
\end{align}
where $\varepsilon_{\mathrm{a}}(k)=2\sqrt{F_{\mathrm{a}}(k)^2-G_{\mathrm{a}}(k)^2}$ with
\begin{align}
G_{\mathrm{a}}(k)&=-J_0\cos^{2}\phi^{\mathrm{c}}_{\mathrm{a}}C_{\alpha}^{(N)}(k),\\
F_{\mathrm{a}}(k)&=G_{\mathrm{a}}(k)-E_{\mathrm{a}}^0,\\
E_{\mathrm{a}}^0&=-2\mathsf{J}^{\mathrm{p}}_{\mathrm{a}}\sin^{2}\phi^{\mathrm{c}}_{\mathrm{a}}-B\cos\phi^{\mathrm{c}}_{\mathrm{a}}.\label{eq.ea0}
\end{align}
Figure~\ref{fig.energyFerroExcited} shows the energy dispersion of spin-wave excitations around the highest excited energy level of a ferromagnet in the thermodynamic limit. Similarly to Fig.~\ref{fig.energyFerro}, the energy dispersion shows a singular behavior, in this case for the mode at $k=\pm\pi$. We conclude that critical behavior, reminiscent of a quantum phase transition, can be observed around the highest excited energy level of a ferromagnet at the critical field
\begin{align}
B_{\mathrm{a}}^{\mathrm{c}}=-2\mathsf{J}^{\mathrm{p}}_{\mathrm{a}},
\end{align}
which, according to Eq.~(\ref{eq.limitJa}), converges to $\lim_{N\rightarrow\infty}B_{\mathrm{a}}^{\mathrm{c}}=-2J_0\eta(\alpha)$.

The considered configuration also reproduces the ground state of an anti-ferromagnet in the case $J_0<0$. Thus, the results of the present Section can be interpreted as the spin wave excitations of the anti-ferromagnetic ground state. In this case, the vanishing excitation gap indicates a ``conventional'' ground-state quantum phase transition. Conversely, the results of Section~\ref{sec.excitationsfmgst} describe the critical behavior of the highest excited state of an anti-ferromagnet when $J_0<0$.

\subsection{Quantum fluctuations around intermediate energy states}
Let us finally illustrate the application of the methods developed in the previous Sections to an example of an intermediate excited state of the system. For simplicity, we consider the absence of a transverse field, i.e., $B=0$. We focus here on the description of the configuration $\boldsymbol{\phi}_{\mathrm{I}}=(\phi+\pi,\phi,\phi+\pi,\dots)$, which is created by performing a spin flip at every odd site of a uniform configuration. Using $\sin\phi^{(i)}=(-1)^{i}\sin\phi$ and $\cos\phi^{(i)}=(-1)^{i}\cos\phi$, it is direct to obtain two sets of extremal points of the energy landscape~\eqref{eq.meanfieldpbc}, characterized by $\cos\phi_{\mathrm{I}_{1}}=0$ and $\sin\phi_{\mathrm{I}_{2}}=0$, respectively.

In the case of the first solution $\phi_{\mathrm{I}_{1}}$, the system's semiclassical energy is given by $E^{\mathrm{p}}(\boldsymbol{\phi}_{\mathrm{I}})|_{\phi=\phi_{\mathrm{I}_{1}}}=-\mathsf{J}_{\mathrm{a}}^{\mathrm{p}}$, and the quadratic Hamiltonian~\eqref{eq.HaBosPBC} has already diagonal form $H^{\mathrm{p}}_{\mathrm{Q}}(\boldsymbol{\phi}_{\mathrm{I}})|_{\phi=\phi_{\mathrm{I}_{1}}}=4\mathsf{J}_{\mathrm{a}}\sum_{i=1}^{N} a_{i}^{\dagger}a_{i}$.

For the second solution $\phi_{\mathrm{I}_{2}}$, we obtain the energy $E^{\mathrm{p}}(\boldsymbol{\phi}_{\mathrm{I}})|_{\phi=\phi_{\mathrm{I}_{2}}}=0$. Correspondingly, the quantum fluctuations are governed by the Hamiltonian 
\begin{align}
            H^{\mathrm{p}}_{\mathrm{Q}}(\boldsymbol{\phi}_{\mathrm{I}})|_{\phi=\phi_{\mathrm{I}_{2}}}=-\frac{J_0}{2}\sum_{r\in I^0_N} \frac{(-1)^{r}}{|r|^\alpha}(a_{i}^{\dagger}+a_{i})(a_{i+r}^{\dagger}+a_{i+r}).
\end{align}
One can map such staggered configuration into a uniform one by means of a unitary transformation $\Pi=\exp\left(\mathrm{i}\pi\sum_{l\:\mathrm{ odd}}a_{l}^{\dagger}a_{l}\right)$. This defines a new Hamiltonian
$\widetilde{H}^{\mathrm{p}}_{\mathrm{Q}}(\boldsymbol{\phi}_{\mathrm{I}})|_{\phi=\phi_{\mathrm{I}_{2}}}=\Pi^{\dagger} H^{\mathrm{p}}_{\mathrm{Q}}(\boldsymbol{\phi}_{\mathrm{I}})|_{\phi=\phi_{\mathrm{I}_{2}}}\Pi$, which reads
\begin{align}
&\quad\widetilde{H}^{\mathrm{p}}_{\mathrm{Q}}(\boldsymbol{\phi}_{\mathrm{I}})|_{\phi=\phi_{\mathrm{I}_{2}}}\notag\\&=-\frac{J_0}{2}\sum_{r\in I^0_N} \frac{1}{|r|^\alpha}(a_{i}^{\dagger}+a_{i})(a_{i+r}^{\dagger}+a_{i+r})
            \nonumber \\&
            =-J_0\sum_{k} C_{\alpha}^{(N)}(k)(A_{k}^{\dagger}A_{-k}^{\dagger}+A_{k}^{\dagger}A_{k}+A_{-k}A_{-k}^{\dagger}+A_{k}A_{-k}).
\end{align}
After a Bogoliubov transformation, one observes that the system exhibits a flat energy dispersion $\varepsilon_{I}(k)=0$. A flat dispersion relation was also
noted for the configurations in the preceding Sections in the absence of a magnetic field, see Figs.~\ref{fig.energyFerro}a) and \ref{fig.energyFerroExcited}a). For the configuration considered here, the rotation $\Pi$ was able to remove the alternating phases, which allowed us to describe the Hamiltonian in terms of a single bosonic species. If this is not the case, a unit cell with several kinds of Bogoliubov bosons is required for the description of fluctuations around intermediate energy states. For completeness, in appendix~\ref{AppendixA} we discuss the case of a lattice with two different types of bosons per unit cell. A careful analysis of the two-species scenario at a saddle-point configuration of the semiclassical energy~(\ref{eq.meanfieldpbc}) is beyond the scope of the present paper. Due to the dynamical instability, such a configuration would lead to a complex energy spectrum and the breakdown of the second-order expansion~(\ref{eq:HamTaylorS}) at long times.

\section{Summary and conclusions}

To summarize, a semiclassical energy landscape was derived by a variational ansatz in terms of spin-coherent states, and, equivalently, as the lowest-order term of a formal $1/\sqrt{2S}$-expansion. Employing a series of single-parameter spin configurations, we produced one-dimensional projections of this $N$-dimensional energy landscape. This provided a simple semiclassical approximation of the full quantum spectrum, which is exact for vanishing or very strong external fields---independently of the length $S$ of the spins, and of the interaction range. As the strength of the external field is increased, each of the mean-field signatures of the semiclassical excited-state levels exhibits a bifurcation. In the case of the ground state, the bifurcation of the energy landscape is directly related to the broken symmetry characterizing the \mbox{(anti-)}ferromagnetic phase.

By studying the spin-wave fluctuations about the semiclassical energy for a selection of spin-configurations, we identified quantum signatures of the bifurcation of the energy landscape in the dispersion relations of spin waves around some of the excited states. Furthermore, in the semiclassical limit $S=M/2\gg 1$, a closing excitation gap of the elementary excitations around the ground state predicts the exact magnetic field at which a quantum bifurcation occurs for the Hamiltonian~(\ref{eq.spinchainS}) of an arbitrary number $N$ of long spins with tunable-range interactions.

Interestingly, our model allows us to explore the semiclassical and the thermodynamic limits separately and in a controlled fashion. For a finite number of sites $N$, the semiclassical limit is reached by allowing for a large number $M$ of non-interacting elementary spins at each site. The thermodynamic limit, in turn, is approached by increasing the number $N$ of interacting, composite spins. 
For finite $S$, we observe a quantum phase transition in the thermodynamic limit, which, e.g., includes the well-known transition of the Ising model for $S=1/2$ and $\alpha=\infty$. This quantum phase transition occurs among the $N\gg 1$ \textit{composite} spins of length $S$, whose interaction range is characterized by $\alpha$. For finite interaction ranges ($\alpha>0$), semiclassical mean-field methods typically fail to make quantitative predictions close to criticality. Conversely, in the semiclassical limit the $M\gg 1$ \textit{elementary} spins, which are always concentrated in a single point, can be represented by few effective degrees of freedom, whose spectral spread depends on $M$. Consequently, in the semiclassical limit, the mean-field prediction becomes exact for all values of $N$ and $\alpha$. The observed non-analytic phenomena are referred to as quantum bifurcations, since they are direct quantum signatures of the bifurcation of the semiclassical energy landscape. 
The long-range interacting spin-$S$ model with transverse field, Eq.~(\ref{eq.spinchainS}), thus provides a family of Hamiltonians which exhibit both quantum bifurcations and quantum phase transitions. Previously studied phase transitions in models with infinite connectivity \cite{ESQPT2}, such as the Lipkin-Meshkov-Glick \cite{Botet} or the Dicke model \cite{Brandes03}, according to the definition presented here, are to be identified as quantum bifurcations rather than as quantum phase transitions.

One advantage of the semiclassical limit $S=M/2\gg 1$ is that for a small number $N$ of sites, one could perform a complete semiclassical analysis in terms of trajectories in phase space. For example, for a system with $N=4$, one could find all the critical points of the energy landscape of Eq.~\eqref{eq:HamTaylorEnergy}. In this case, the existence of saddle points would lead to singularities in derivatives of the density of states, which are referred to as excited-state quantum phase transitions~\cite{ESQPT, ESQPT2}. In this context, it would be interesting to explore the character of the singularities of the density of states in the thermodynamic limit $N\rightarrow\infty$ \cite{Kastner} by considering a description in terms of field theory~\cite{Sachdev}. In future work, it will be interesting to explore the dynamical consequences of the geometry of the energy landscape, i.e., the evolution of the quantum correlations
 \cite{nacho2004} when the system is initially prepared in a coherent state centered at an unstable fixed point. Another possibility is to study the effect 
of an external driving, which enables control of the geometry of quasienergy landscapes~\cite{Victor,abu+dod97}.

\acknowledgments
V. M. Bastidas acknowledges valuable discussions with S. Vinjanampathy. V. M. Bastidas and T. Brandes acknowledge financial support from DFG in the framework of SFB 910. M. Gessner thanks the German National Academic Foundation for support.

\appendix

\section{Bosonized Hamiltonian with periodic boundary conditions}\label{AppendixB}
To derive the bosonized expansion of the spin Hamiltonian~(\ref{eq.spinHpbc}) with periodic boundary conditions in the thermodynamic limit, we follow the procedure introduced in Section~\ref{sec.bosons}. Employing the rotation $U(\boldsymbol{\phi})$, as introduced below Eq.~(\ref{eq:UnitCoherentState}), followed by a Holstein-Primakoff bosonization~\cite{HP,Dusuel}, we obtain the following expression for $H^{\mathrm{p}}(\boldsymbol{\phi})=U(\boldsymbol{\phi})H^{\mathrm{p}}U^{\dagger}(\boldsymbol{\phi})$:


\begin{align}\label{eq.hphiappendix}
H^{\mathrm{p}}(\boldsymbol{\phi})&=-SJ_0\sum_{i=1}^{N}\sum_{r\in I^0_N}\frac{\sin\phi^{(i)}\sin\phi^{(i+r)}}{|r|^{\alpha}}-2SB\sum_{i=1}^{N}\cos\phi^{(i)}\notag\\
&\quad+J_0\sqrt{2S}\sum_{i=1}^{N}\sum_{r\in I^0_N}\frac{(a_i+a_i^{\dagger})\cos\phi^{(i)}\sin\phi^{(i+r)}}{|r|^{\alpha}}\notag\\
&\quad-B\sqrt{2S}\sum_{i=1}^{N}(a_i+a_i^{\dagger})\sin\phi^{(i)}\notag\\
&\quad+2J_0\sum_{i=1}^{N}\sum_{r\in I^0_N}\frac{a_{i}^{\dagger}a_{i}\sin\phi^{(i)}\sin\phi^{(i+r)}}{|r|^{\alpha}}\notag\\
&\quad-\frac{J_0}{2}\sum_{i=1}^{N}\sum_{r\in I^0_N}\frac{\cos\phi^{(i)}\cos\phi^{(i+r)}(a_i+a_i^{\dagger})(a_{i+r}+a_{i+r}^{\dagger})}{|r|^{\alpha}}\notag\\
&\quad+2B\sum_{i=1}^{N}a_i^{\dagger}a_i\cos\phi^{(i)}.
\end{align}
We introduce the mean-field energy for periodic boundary conditions
\begin{align}
E^{\mathrm{p}}(\boldsymbol{\phi})=-\frac{J_0}{2N}\sum_{i=1}^{N}\sum_{r\in I^0_N}\frac{\sin\phi^{(i)}\sin\phi^{(i+r)}}{|r|^{\alpha}}-\frac{B}{N}\sum_{i=1}^{N}\cos\phi^{(i)}
\end{align}
with its first derivatives
\begin{align}\label{eq.pbcfirstder}
\frac{\partial E^{\mathrm{p}}(\boldsymbol{\phi})}{\partial \phi^{(i)}}=-\frac{J_0}{N}\sum_{r\in I^0_N}\frac{\cos\phi^{(i)}\sin\phi^{(i+r)}}{|r|^{\alpha}}+\frac{B}{N}\sin\phi^{(i)},
\end{align}
and second derivatives
\begin{align}\label{eq.pbcsecondderdiag}
\frac{\partial^2 E^{\mathrm{p}}(\boldsymbol{\phi})}{\partial \phi^{(i)2}}=\frac{J_0}{N}\sum_{r\in I^0_N}\frac{\sin\phi^{(i)}\sin\phi^{(i+r)}}{|r|^{\alpha}}+\frac{B}{N}\cos\phi^{(i)},
\end{align}
and
\begin{align}
\frac{\partial^2 E^{\mathrm{p}}(\boldsymbol{\phi})}{\partial \phi^{(i)}\partial \phi^{(i+r)}}=-\frac{J_0}{N}\frac{\cos\phi^{(i)}\cos\phi^{(i+r)}}{|r|^{\alpha}}.
\end{align}
This allows us to reexpress Eq.~(\ref{eq.hphiappendix}) as
\begin{align}\label{eq.expansionp}
H^{\mathrm{p}}(\boldsymbol{\phi})&=2SNE^{\mathrm{p}}(\boldsymbol{\phi})+\sqrt{2S}H^{\mathrm{p}}_{\mathrm{L}}(\boldsymbol{\phi})+H^{\mathrm{p}}_{\mathrm{Q}}(\boldsymbol{\phi})+\mathcal{O}\left(\frac{1}{\sqrt{2S}}\right)
\end{align}
where we have introduced
\begin{align}\label{eq.HlinA}
H^{\mathrm{p}}_{\mathrm{L}}(\boldsymbol{\phi})=-N\sum_{i=1}^{N}\frac{\partial E^{\mathrm{p}}(\boldsymbol{\phi})}{\partial \phi^{(i)}}(a_i+a_i^{\dagger}),
\end{align}
and
\begin{align}\label{eq.HquadA}
H^{\mathrm{p}}_{\mathrm{Q}}(\boldsymbol{\phi})&=\frac{N}{2}\sum_{i=1}^{N}\sum_{r\in I^0_N}\frac{\partial^2 E^{\mathrm{p}}(\boldsymbol{\phi})}{\partial \phi^{(i)}\partial \phi^{(i+r)}}(a_i+a_i^{\dagger})(a_{i+r}+a_{i+r}^{\dagger})\notag\\
&\quad+2N\sum_{i=1}^{N}\frac{\partial^2 E^{\mathrm{p}}(\boldsymbol{\phi})}{\partial \phi^{(i)2}}a_i^{\dagger}a_i.
\end{align}
Note also that, given $\partial E^{\mathrm{p}}(\boldsymbol{\phi})/\partial \phi^{(i)}=0$ and $\cos\phi^{(i)}\neq 0 \:\forall i$, Eq.~(\ref{eq.pbcsecondderdiag}) simplifies to
\begin{align}
\frac{\partial^2 E^{\mathrm{p}}(\boldsymbol{\phi})}{\partial \phi^{(i)2}}=\frac{B}{N}\frac{1}{\cos\phi^{(i)}},
\end{align}
in analogy to Eq.~(\ref{eq.simpldiagsecder}). If the above conditions are satisfied, this expression may be used to simplify the quadratic Hamiltonians at extremal points, by modifying for instance the respective second terms of Eqs.~(\ref{eq:HamTaylorQuadGroundFerr}), (\ref{eq:HamTaylorQuadGroundRecipFerr}) and (\ref{eq:HamTaylorQuadStaggered}), as well as expressions~(\ref{eq.ee0}) and (\ref{eq.ea0}).

\section{Quantum fluctuations in the case of two sublattices\label{AppendixA}}

Similarly to Ref.~\cite{Yosida}, let us consider an index $l=1,2\ldots,N/2$ to label the unit cell and a partition of the system into two sublattices $B$ and $C$ with lattice vectors $i_{\mathrm{B}}(l)=2l-1$ and $i_{\mathrm{C}}(l)=2l$, respectively. For simplicitly, we restrict to the case an even number $N$ of spins. The two sublattices are characterized by angles $\phi_{B}$ and $\phi_{C}$ though the configuration $\boldsymbol{\phi}_{\mathrm{BC}}=(\phi_{B},\phi_{C},\ldots,\phi_{B},\phi_{C})$. In this case, the condition Eq.~\eqref{eq.pbcfirstder} reads $\partial E^{\mathrm{p}}(\boldsymbol{\phi})/\partial \phi_{B}=\partial E^{\mathrm{p}}(\boldsymbol{\phi})/\partial \phi_{C}=0$, which is equivalent to the two coupled equations
\begin{align}
      \label{eq:ExtSubLatt}
             -J_{0}\cos\phi_B(M^{(N)}_{B}\sin\phi_C+M^{(N)}_{C}\sin\phi_{B})+B\sin\phi_B&=0
             \nonumber\\
             -J_{0}\cos\phi_C(M^{(N)}_{B}\sin\phi_{B}+M^{(N)}_{C}\sin\phi_C)+B\sin\phi_C&=0
      \ ,
\end{align}
where 
\begin{align}
M^{(N)}_{B}&=\sum_{\substack{r\in I^0_N\\r\:\mathrm{odd}}}\frac{1}{|r|^{\alpha}},\\
\intertext{and}
M^{(N)}_{C}&=\sum_{\substack{r\in I^0_N\\r\:\mathrm{even}}}\frac{1}{|r|^{\alpha}}.
\end{align}
The set $I_N^0$ was introduced below Eq.~(\ref{eq.spinHpbc}). In addition, we define the operator $(\boldsymbol{\hat{A}}^{\dagger}_{l})^{T}=(a^{\dagger}_{2l-1},a^{\dagger}_{2l})$.
Similarly, we define $\boldsymbol{\hat{X}}_{l}=\boldsymbol{\hat{A}}_{l}+\boldsymbol{\hat{A}}^{\dagger}_{l}$. By using these definitions, we can write the Hamiltonian~\eqref{eq.HaBosPBC} for the configuration $\boldsymbol{\phi}_{\mathrm{BC}}$ in a simple way:
\begin{align}
      \label{eq:HamTaylorQuadGroundNonTriv}
            H^{\mathrm{p}}_{\mathrm{Q}}(\boldsymbol{\phi}_{\mathrm{BC}})&=\sum^{N/2}_{l=1}(\boldsymbol{\hat{A}}^{\dagger}_{l})^{\mathrm{T}}\boldsymbol{\mathcal{M}}^{(N)}\boldsymbol{\hat{A}}_{l}+\sum^{N/2}_{l=1}(\boldsymbol{\hat{X}}^{\dagger}_{l})^{\mathrm{T}}\boldsymbol{\mathcal{N}}\boldsymbol{\hat{X}}_{l}
              \\ &\quad
            +\sum_{l}^{N/2}\sum_{R\in I_{N/2}^0}\left[(\boldsymbol{\hat{X}}^{\dagger}_{l})^{\mathrm{T}}\boldsymbol{\mathcal{K}}_{|R|}\boldsymbol{\hat{X}}_{l+R}\right]\notag,
\end{align}
where 
\begin{align}
       \label{eq:LocalFreq}
       \boldsymbol{\mathcal{M}}^{(N)} =\left(%
\begin{array}{ccc}
\omega^{(N)}_{B}  &  0  \\
 0 & \omega^{(N)}_{C}
\end{array}
\right)
\end{align}
and 
\begin{align}
      \label{eq:Frequency}
             \omega^{(N)}_{B,C}&=J_0\sin\phi_{B,C}(M^{(N)}_{B}\sin\phi_{C,B}+M^{(N)}_{C}\sin\phi_{B,C})
             \nonumber\\&\quad
             +2B\cos\phi_{B,C}.
\end{align}

In a similar way, we describe the coupling of the Bogoliubov bosons within the $l$-th unit cell by using the matrix
\begin{align}
       \boldsymbol{\mathcal{N}} =-\frac{J_0}{2}\cos\phi_{B}\cos\phi_{C}\left(%
\begin{array}{ccc}
0  &  1  \\
 1 & 0 
\end{array}
\right)
\end{align}
and the inter-cell coupling matrix
\begin{align}
       \label{eq:NonLocalMatOsciSqu}
       \boldsymbol{\mathcal{K}}_{R} =-J_{0}\left(%
\begin{array}{ccc}
\frac{\cos\phi^{2}_{B}}{(2R)^{\alpha}} &  \frac{\cos\phi_{B}\cos\phi_{C}}{(2R+1)^{\alpha}} \\
 \frac{\cos\phi_{B}\cos\phi_{C}}{(2R-1)^{\alpha}} & \frac{\cos\phi^{2}_{C}}{(2R)^{\alpha}}
\end{array}
\right)
\ .
\end{align}
In a similar way to the discussion of the ferromagnetic case, we introduce here a discrete Fourier transformation
\begin{align}
	\boldsymbol{\hat{\mathcal{A}}}_{l}=\sqrt{\frac{2}{N}}\sum_{k}\boldsymbol{\hat{A}}_{k} e^{\mathrm{i} k l}
	\label{eq:operatorDFT}
	\ .
\end{align}
In addition, one can show that in the particle-hole basis, $(\boldsymbol{\hat{\Psi}}^{\dagger}_{k})^{T}=(\boldsymbol{\hat{\mathcal{A}}}^{\dagger}_{k},\boldsymbol{\hat{\mathcal{A}}}_{-k})$, one can write the Hamiltonian~\eqref{eq:HamTaylorQuadGroundNonTriv} as $H^{\mathrm{p}}_{\mathrm{Q}}(\boldsymbol{\phi}_{\mathrm{BC}})=\sum_{k}(\boldsymbol{\hat{\Psi}}^{\dagger}_{k})^{T}\boldsymbol{H}_{k}\boldsymbol{\hat{\Psi}}_{k}$. Correspondingly, we define the Bogoliubov de Gennes Hamiltonian
\begin{align}
       \label{eq:LocalMatOsciSqu}
       \boldsymbol{H}_{k} =\left(%
\begin{array}{ccc}
\boldsymbol{\mathcal{M}}^{(N)}+2\boldsymbol{\mathcal{N}}+2\mathrm{Re}(\boldsymbol{\mathcal{K}}_{k})  &  2\boldsymbol{\mathcal{N}}+2\mathrm{Re}(\boldsymbol{\mathcal{K}}_{k})  \\
 2\boldsymbol{\mathcal{N}}+2\mathrm{Re}(\boldsymbol{\mathcal{K}}_{k}) & \boldsymbol{\mathcal{M}}^{(N)}+2\boldsymbol{\mathcal{N}}+2\mathrm{Re}(\boldsymbol{\mathcal{K}}_{k}) 
\end{array}
\right) .
\end{align}

In the thermodynamic limit, one obtains the expressions $\lim_{N\rightarrow\infty}M^{(N)}_{B}=2(1-2^{-\alpha})\zeta(\alpha)$ and $\lim_{N\rightarrow\infty}M^{(N)}_{C}=2^{-\alpha+1}\zeta(\alpha)$, which enable one to calculate the thermodynamic limit of the matrix $\boldsymbol{\mathcal{M}}^{(N)}$.
Interestingly, in this limit, the effect of the long-range interactions between the bosonic particles is included in the Fourier transformation of the coupling matrix Eq.~\eqref{eq:NonLocalMatOsciSqu}, which reads
\begin{align}
       \boldsymbol{\mathcal{K}}_{k} =\frac{-J_{0}}{2^{\alpha}}\left(%
\begin{array}{ccc}
\zeta(\alpha)\cos\phi^{2}_{B}  &  L_{3/2}(k)\cos\phi_{B}\cos\phi_{C}  \\
L_{1/2}(k)\cos\phi_{B}\cos\phi_{C} & \zeta(\alpha)\cos\phi^{2}_{C} 
\end{array}
\right) ,
\end{align}
where $L_{1/2}(k)=\Phi(e^{\mathrm{i}k},\alpha,1/2)$, $L_{3/2}(k)=\Phi(e^{\mathrm{i}k},\alpha,3/2)$, and $\Phi(\lambda,n,a)$ is the Lerch transcendent function~\cite{AndrewsLerch, AbramowitzStegun}.


\begin{thebibliography}{xx}
\bibitem{footnote2} For the continuous parametrization of energy levels we do not consider those configurations which lead to the eigenvalue zero in either of the two limits $B=0$ or $J_0=0$, since they would generate unphysical solutions at the respective other end of the spectrum.
\end{thebibliography}
\end{document}